\documentclass[twocolumn,10pt,a4paper]{article}
\pdfoutput=1
\usepackage{graphicx,subfigure}
\usepackage{amsmath,amssymb,amsfonts}
\usepackage{german}
\usepackage{placeins}

\begin{document}
\sffamily
\newcommand{\engl}{\selectlanguage{english}}
\newcommand{\germ}{\selectlanguage{german}}
\engl
\title{\sf Long Range Structure of the Nucleon}
\author{Marc Vanderhaeghen and Thomas Walcher\\
\large Institut f\"ur Kernphysik\\ Johannes-Gutenberg-Universit\"at Mainz}
\maketitle

\begin{abstract}
The long range structure of the nucleon is discussed starting from the old 
model of a quark bag with a pion cloud (``cloudy bag'') carrying on to the 
more recent ideas of the parton model of the nucleon. On the basis of the 
most recent measurements of the form factors at MAMI, JLab and MIT quantitative
results for nucleon charge densities are presented within both non-relativistic 
and relativistic frameworks.    
\end{abstract}


\section{\sf Introduction: Ranges }

Many physicists have the following picture of the structure of the nucleon : In
the inner region at ``short range'' reside quarks bound by gluons, in the 
outer region at ``large distances'' live mesons and in particular pions. 
The nucleon consists of constituents, i.e. ``constituent quarks'' and pions,   
as the atom consists of a nucleus and electrons, and the nucleus of protons and 
neutrons. The range of the inner ``coloured'' region, frequently called 
``confinement radius'', is rather elusive. It is a model parameter  
in the old bag model or models for the nucleon resonances based on the 
constituent quarks. One can estimate it from experiment by 
identifying it with the ``annihilation radius'' of the antiproton-proton 
system. Only quark-antiquark annihilation in the overlap region of colour 
can contribute to annihilation~\cite{Povh:1986uk}. It amounts to 
approximately 0.8\,fm, in reasonable agreement with the mentioned model 
parameters. However, it cannot be easily identified with the 
root-\-mean-\-square (rms) radius of the electric charge of the proton,  
since the pion cloud will contribute to the charge distribution. A rough 
idea of the range of this contribution may be gotten by the Compton wave 
length of the pion which is of the order of 1.4\,fm. A similar picture emerges 
from diffractive scattering of high energy protons and 
antiprotons~\cite{Islam:2010}. 

However, this picture fails in two ways. Firstly, the nucleon moves after 
the scattering in most experiments at relativistic velocities and therefore 
its structure looks different in different reference frames. The simplest 
example is the transformation of the magnetic moment into an electric dipole 
moment. This situation makes it particularly difficult to compare experiments 
to model calculations based on rest frame wave functions. These wave functions 
have to be ``boosted'' to the correct momentum transfer and there is no 
consistent way of doing that. Secondly, at relativistic energies 
particle-\-antiparticle, i.e. quark-\-antiquark, pairs have unavoidably 
to be considered making the picture much more involved. These two aspects 
will be discussed in Section\,\ref{sec:3}. 

The relativity destroys the simple picture also in another way. We usually
relate ranges with momentum transfer via Heisenberg's uncertainty
relation. However, since in relativistic mechanics space and time are 
intimately connected no relativistic uncertainty relation 
exists~\cite{LanL:1971}. Therefore the assignment of ranges to quantities 
depending on the negative four momentum transfer squared $Q^2=-q^2$, 
as e.g. in some plots of the running coupling constant, is rather 
misleading. We shall, therefore, in this article distinguish between the 
non-\-relativistic picture derived at small momentum transfers and the 
relativistic case where we have to use the relativistic quantum field theoretic 
description. At small momentum transfers we can approximate the four momentum 
transfer $q$ with the three momentum transfer $\vec{q}^{\,2}\approx -q^2$ and 
maintain the familiar interpretation of form factors (Section\,\ref{sec:2}). 
In Section\,\ref{sec:3} we shall show how we can connect the 
non-\-relativistic picture to the underlying quark-gluon structure. We 
shall see that new experiments in just the relativistic domain are 
needed in order to clarify how nucleons are made up of quarks and gluons or 
more precisely how hadrons emerge from QCD.


\section{\sf Nonrelativistic interpretation of nucleon form factors} 
\label{sec:2}

The non-relativistic electromagnetic form factor (FF) is a special case of one of
the fundamental observables in quantum mechanics. It is the matrix element of
the interaction propagator of the exchange photon, i.e. the operator
$e^{i\vec{q} \cdot \vec{r}}$. It appears in all domains of micro physics
ranging from atomic physics, the M\"ossbauer effect to particle 
physics~\cite{Lip:1973}. It reads in the case of elastic scattering from a charge 
density $\rho(\vec{r})= \langle \vec{r}|\vec{r}\rangle$
\begin{equation}
F(\vec{q}) =  \int d^3 \vec r \, \langle \vec{r}\,|\,e^{i\vec{q} 
              \cdot \vec{r}} \,|\, \vec{r}\rangle.
\label{eq:FF}
\end{equation} 
Since we have the idea that the photon interacts with a single constituent we
can interpret the form factor as the probability that we scatter from a
constituent with a momentum just right to leave the bound system ``intact'',
i.e. in the ground state. Since one deals mostly with spherically symmetric systems 
it is customary to write $F(q^{\,2})$. 

The method to derive charge and magnetic distributions from the scattering of
electro-\-magnetically interacting particles - almost exclusively these are
electrons - has been extensively used for nuclei and nucleons. As a reminder 
we present the Rosenbluth formula, the key formula connecting the measured cross
sections to the FFs for a spin-1/2 particle~:
\begin{multline}
\left. \frac{d\sigma}{d\Omega}\right|_{lab} = 
\underbrace{\left( \frac{\alpha^2}{4E^2
sin^4(\theta/2)} \right)}_{\textstyle \sigma_{\text{Mott}}}
\frac{E'}{E} 
\\
\times \biggl\{ 
\frac{G^2_E + \tau G^2_M}{1+\tau}\, cos^2 (\theta/2) 
  - 2 \tau G^2_M \, sin^2(\theta/2) \biggr\}
\label{eq:Rosen}  
\end{multline}
where $E$ is the incoming and $E'$ the outgoing electron energy,
$\theta$ the scattering angle, and $\tau = Q^2/(4 M^2)$ with $M$ 
the mass of the nucleon. Dividing the measured cross section by the 
Mott cross section, i.e. the cross section for point particles, 
yields the combination of the quadratic electric form factor $G^2_E$ 
and magnetic $G^2_M$. By varying the scattering angle $\theta$ and  
the incoming electron energy $E$ in such a way that $Q^2$ is constant, 
$G^2_E$ and $G^2_M$ can be separated, the so called Rosenbluth 
separation. This formula is relativistically correct meaning that 
the FFs are functions of $Q^2$. As a practical tool, the Rosenbluth 
method is not ideal, however, for very small cross sections as is the 
case when extracting the small $G^2_E$ term of the neutron, or when 
separating at large $Q^2$ the small $G^2_E$ contribution of the proton 
from that of $G^2_M$ since the latter dominates at large momentum 
transfers. Here polarized electrons available with sufficient intensity 
since about a decade, together with polarized targets or recoil polarimetry, 
have changed our possibilities dramatically. 

\begin{figure}[h]
\begin{center}
\includegraphics[width=0.80\columnwidth]{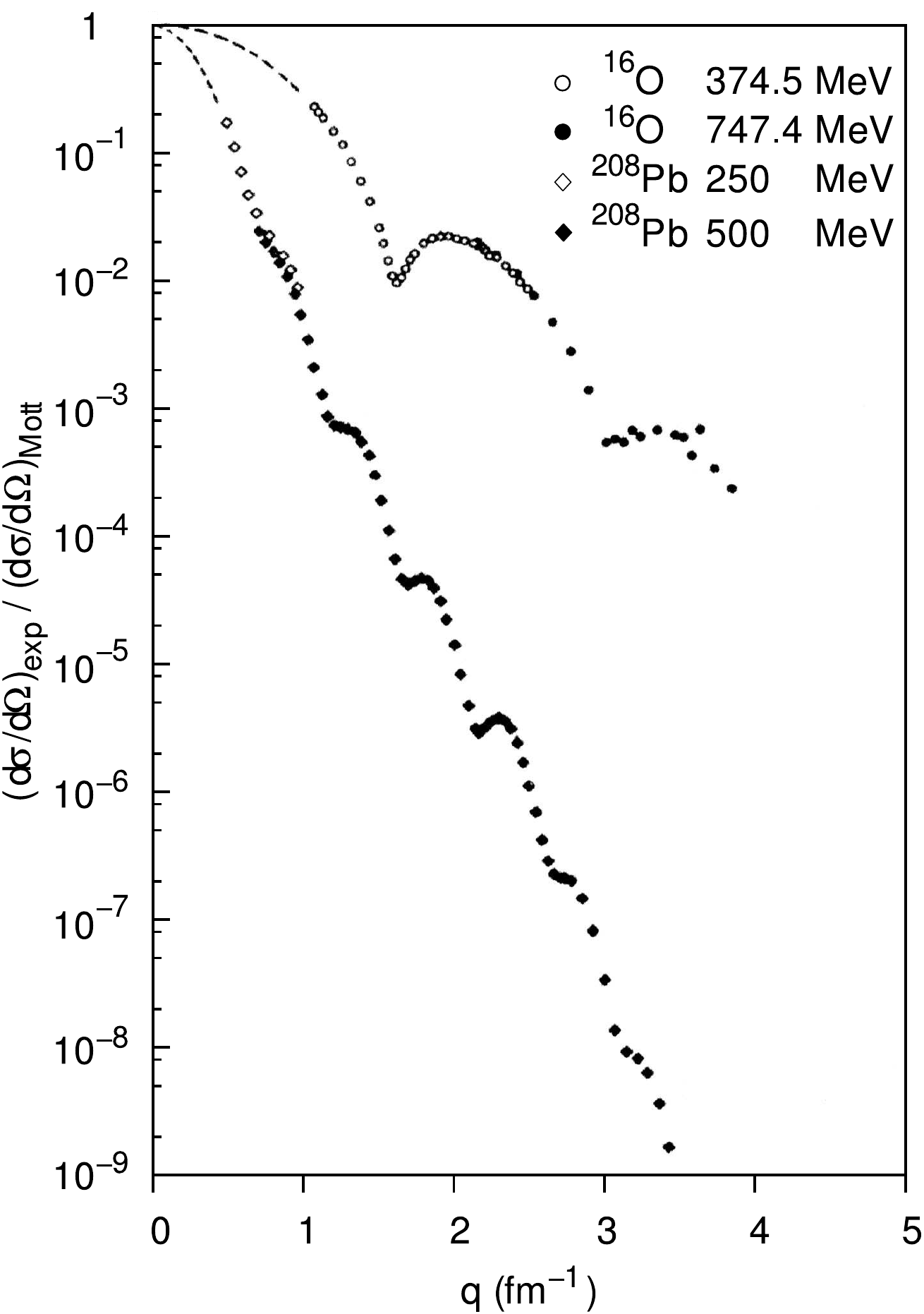} 
\end{center}
\caption{A compilation of form factors (FFs) from elastic electron scattering 
  for $^{16}O$ and $^{208}Pb$ measured at the electron-\-accelerator 
  laboratories: Amsterdam, Darmstadt, Mainz, Saclay, and Stanford. The 
  FFs have been corrected for the Coulomb distortions different 
  at the electron energies available at these laboratories. The diffraction 
  pattern determining the charge distribution is nicely seen. 
  (From Ref.\,\cite{Fri:1982}).}
\label{fig:1}
\end{figure}

In order to give a feeling for the significance of the method we present
the example of nuclei. Nuclei are heavy and have in many cases no magnetic
contribution, hence the charge distribution can be precisely determined. 
Figure\,\ref{fig:1} shows a collection of data for the scattering of 
electrons from $^{16}O$ and $^{208}Pb$. 

\begin{figure}[h]
\begin{center}
\includegraphics[width=0.95\columnwidth]{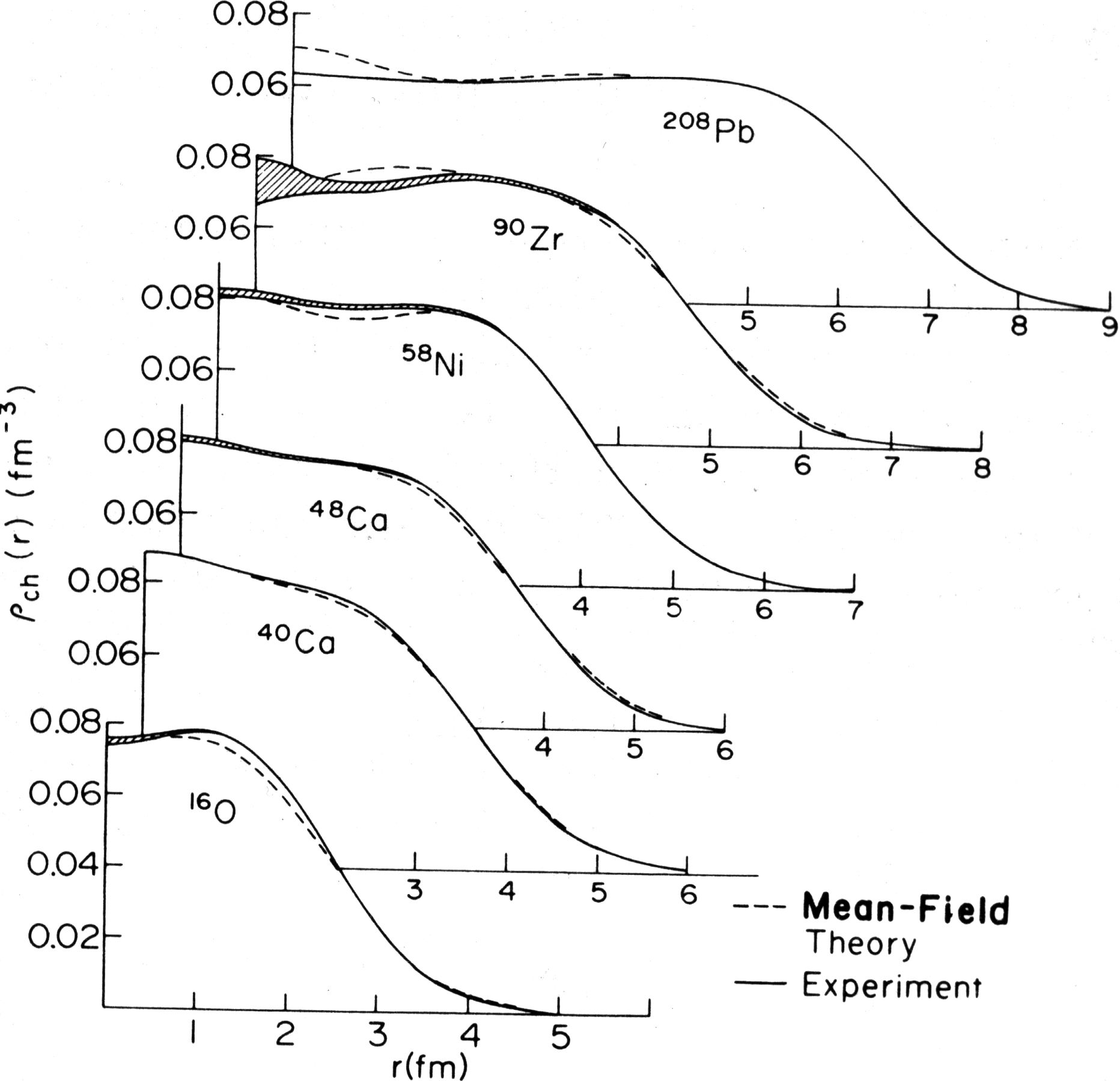}
\end{center}
\caption{Charge distributions of spherical nuclei as derived by model
  independent analyses from experiment (solid lines). The hatched areas
  indicate the range of the charge consistent with the data. The dashed 
  lines depict the results of calculations in the framework of nuclear 
  mean-field theory. (From Ref.\,\cite{Neg:1982})}
\label{fig:2}
\end{figure}

The Fourier transform (see Eq.\,(\ref{eq:FF})) of the FFs derived from 
these cross sections allows for a model independent precise determination of 
the charge distributions. This is shown in Fig.\,\ref{fig:2} for some nuclei. 
These distributions can be well modeled by mean field calculations based on 
the nuclear shell model~\cite{Neg:1982}. 
We see that the nuclear charge distributions have a subtle structure on 
top on the bulk behaviour of the saturated nuclear matter. This structure 
reflects the shell model wave functions and is very sensitive to the details 
of the nucleon-nucleon interaction but even more importantly to the many body
features of the nuclei made up of the constituents protons and neutrons. 
Only the most sophisticated mean field theories are able to almost reproduce 
this structure~\cite{Neg:1982}. It remains an unexplained deviation which may 
be due to too narrow error bands~\cite{Ber:2010}. 

Let us now turn to the nucleons and see what we know here (see {\it e.g.}  
Refs.\,\cite{HydeWright:2004gh,Arrington:2006zm,Perdrisat:2006hj} 
for some recent reviews on nucleon form factors).  The pre-2000 data suggest 
that the magnetic and electric form factor of the
proton follow a universal form, the ``dipole form'' 
$G_{D}(Q^2) = 1/(1+ Q^2/\Lambda_{D}^2)^2$, with the scale parameter 
$\Lambda_{D} = 0.843$\,GeV approximately equal to the mass $m_{\rho}$ of the 
$\rho$ meson. The case was closed and considered to be text book material. 
This finding was the basis for the much discussed ``vector dominance model'' in 
strong interactions. (For a recent discussion see ref.\,\cite{Crawford2010:mit}.) 
The $\rho$ meson mass defines a ``small range'' of an exponential distribution, 
somewhat unphysical due to its discontinuity at $r=0$. It was believed that the 
dipole form would also describe the long ranges characterized by the rms radius 
of the proton. A measurement at the High Energy Physics Laboratory HEPL at 
Stanford gave $\sqrt{\langle r^2 \rangle} = (0.805 \pm 0.011)$\,fm 
$\approx \sqrt{12} /m_{\rho}$~\cite{Han:1963}. However, as it is now 
clear, a measurement at Mainz in 1980~\cite{Sim:1980} which yielded 
$\sqrt{\langle r^2 \rangle} = (0.862 \pm 0.012)$\,fm was closer to the 
best value $(0.882 \pm 0.010)$\,fm available today. (For a more detailed 
discussion see Ref.\,\cite{Drechsel:2007sq}.) This value is confirmed by a 
recent high precision determination at the Mainz Microtron (MAMI) 
yielding~\cite{Ber:2010}:
\begin{multline}
\sqrt{\langle r^2 \rangle} = (0.879 \pm 0.005_{\mathrm{stat.}}\pm \\ 
 0.004_{\mathrm{syst.}} \pm 0.005_{\mathrm{model}})\, \mathrm{fm}.
\label{eq:re}
\end{multline}
As discussed in Ref.\,\cite{Drechsel:2007sq} this value is 
significantly larger than the largest value derived from dispersion relations 
and is unexplained in most nucleon models. On the other hand, a recent study 
of the Lamb shift in muonic hydrogen at the Paul-Scherrer Institute in 
Zuerich yielded a very precise value as low as $(0.840 \pm 0.001)$\,fm 
\cite{Pohl:2010zz} in agreement with the upper bound of the dispersion-\-relations 
calculations. It is not credible to assign this five standard deviation 
difference to a deficiency in the theory of the electronic experiments which 
is the same Quantum Electrodynamics as for the muonic hydrogen. However, the 
calculations of the Lamb shift in muonic hydrogen are difficult due to the 
strong distortions of the muon wave function calling for further theoretical
work.

Considering the focus on ``long ranges'' the question arises how much of 
the rms radius is possibly due to a pion cloud. 
There were weak indications that the pion cloud could be directly seen 
in the FFs in a similar way as the shell structure of the nucleus.
This indication came from a coherent analysis of the data available until 
2003 for the electric and magnetic FFs of the proton and 
neutron~\cite{Friedrich:2003iz}. Since 2003 the data base has 
improved so much that we want to base this discussion on the most recent 
measurements.   

Figure\,\ref{fig:3} shows the electric FF of the proton as derived 
from a direct fit of a FF model to data obtained with the 3-Spectrometer 
set-up at MAMI. This method takes advantage of modern computers and fits 
Eq.\,({\ref{eq:Rosen}) to a large set of angular distributions measured at five 
energies 180, 370, 450, 720, and 850\,MeV. All together about 1400 settings 
were measured. In this way the ``measurement at constant $Q^2$'', i.e. the old 
Rosenbluth separation, becomes obsolete and a very broad kinematic range can 
be covered indeed. However, since this method is somewhat unfamiliar 
Fig.\,\ref{fig:4} shows the best fit curve of Fig.\,\ref{fig:3} together with 
the results at those kinematics at which a traditional Rosenbluth separation 
could be done. The agreement is very good.  A point of concern may be the 
analytical model used for the electric form factor $G_{E p}$ and magnetic form 
factor $G_{M p}$ in the direct FF fits. Here about a dozen different forms 
have been used all yielding essentially the same results~\cite{Ber:2010}. 
The rms radius is, however, somewhat dependent on them and the second 
systematic error in Eq.\,(\ref{eq:re}) reflects this dependence.   
   
\begin{figure}[h]
\begin{center}
\includegraphics[width=\columnwidth]{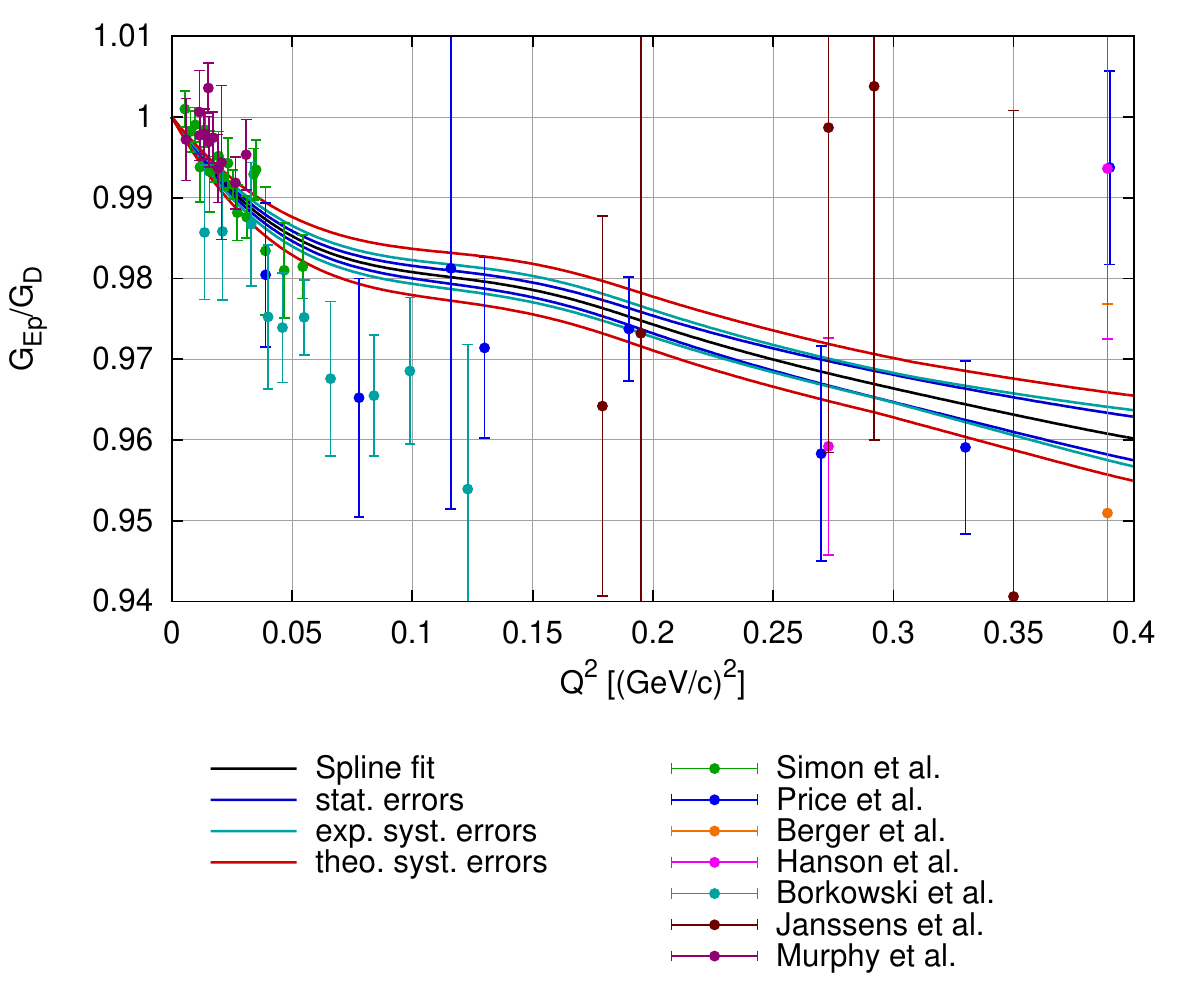}
\end{center}
\caption{The electric FF of the proton $G_{Ep}/G_D$ obtained
from a direct fit to the cross sections with the spline
model to data measured with the 3-Spectrometer set-up at MAMI. 
The FF is normalised to the dipole form given in the text. 
The 1\,$\sigma$-error band is shown for the indicated errors. 
In order to show the improvement, the old data obtained with the 
classical Rosenbluth separation are also depicted.}
\label{fig:3}
\end{figure}

\begin{figure}[h]
\begin{center}
\includegraphics[width=\columnwidth]{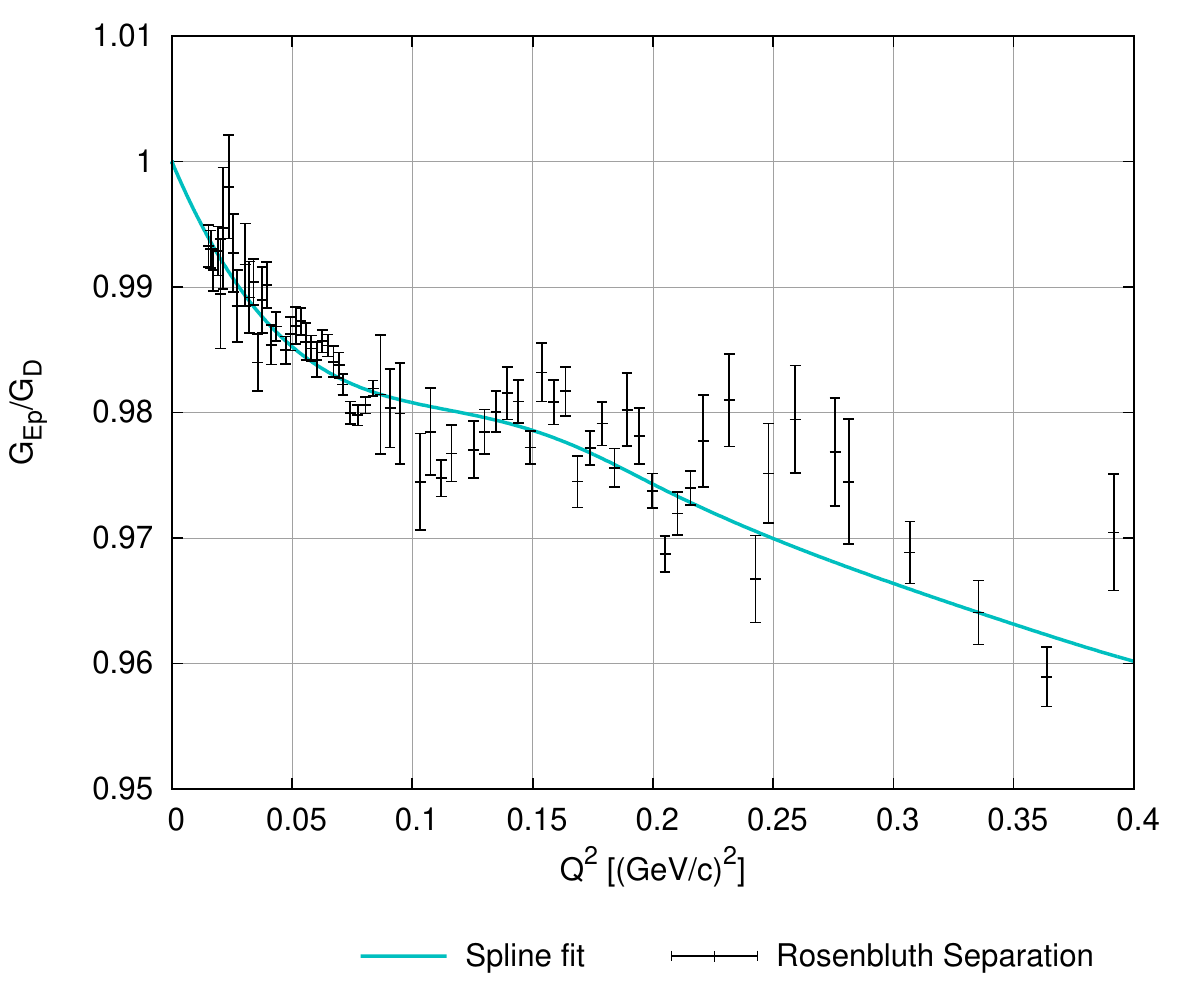}
\end{center}
\caption{As Fig.\,\ref{fig:3} without error bands but with the results from 
the classical Rosenbluth separation of the new MAMI data.}
\label{fig:4}
\end{figure}

\begin{figure}[h]
\begin{center}
\includegraphics[width=\columnwidth]{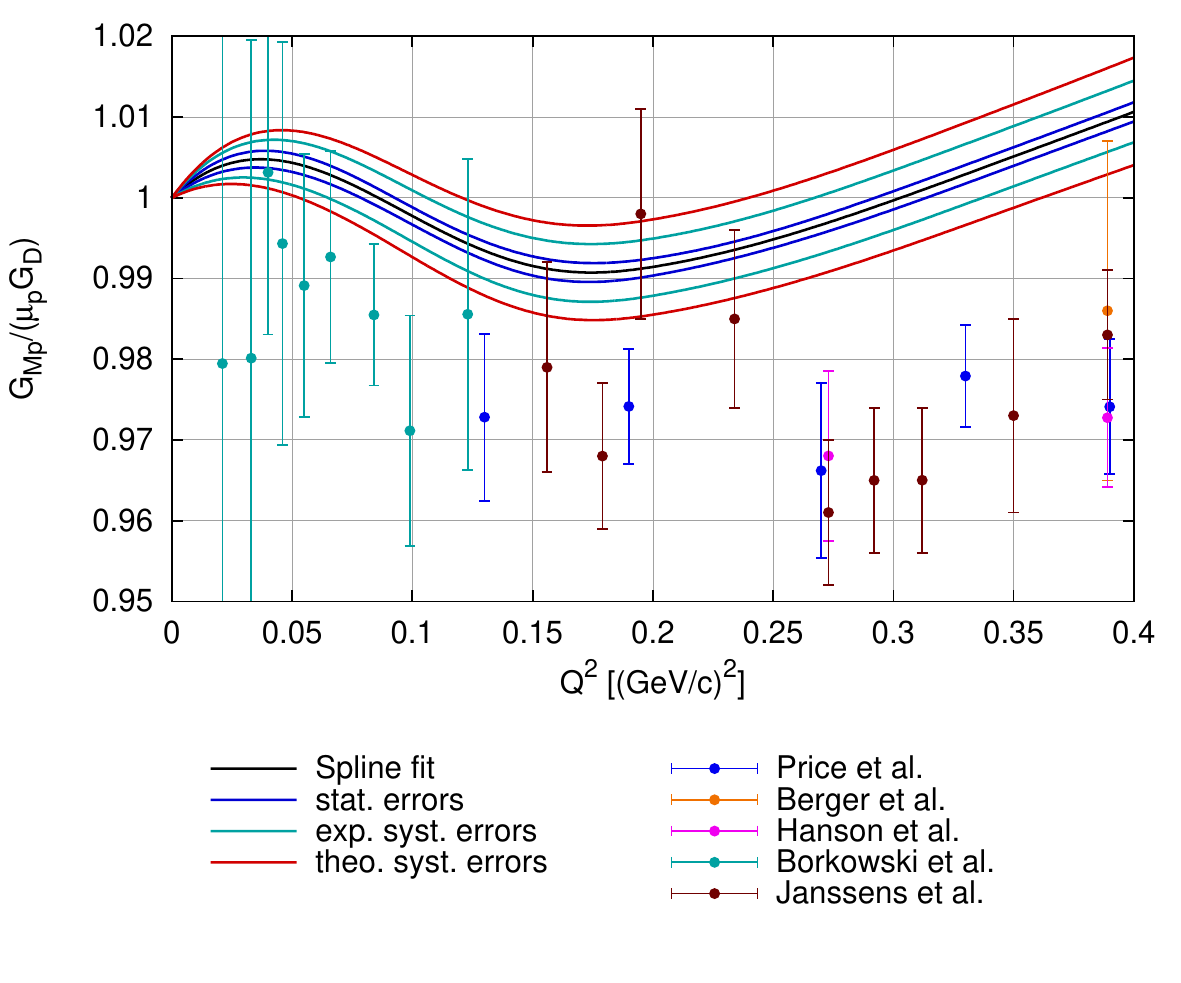}
\end{center}
\caption{As Fig.\,\ref{fig:3} for the magnetic FF   
$G_{Mp}/(\mu_p G_D)$ of the proton.}
\label{fig:5}
\end{figure}

It is evident that these form factors show some structure after the gross 
dependence, assumed to be given by the standard dipole form, has been divided 
out. In Fig.\,\ref{fig:3} one observes two slopes for $G_E/G_D$. The steep 
negative slope at small $Q^2$ is reflected in the large 
rms radius discussed above. The reverse is true in Fig.\,\ref{fig:5} for the rms 
radius of $G_M/(\mu_p G_D)$. A shoulder structure is indicated in both 
FFs at $Q^2 \approx 0.15\,\mathrm{GeV}^2$. It is shifted compared to 
the bump structure derived by Friedrich and Walcher~\cite{Friedrich:2003iz} 
from the pre-2003 data and cannot be identified with it. However, just 
considering the scale of the rms radius and the scale of the structure it is 
suggestive to look for pion cloud contributions in the modelling of the 
nucleon. 

Since the bulk charge of the proton resides at ``small ranges'' and extends 
out to the range of the pion cloud, the separation of the inner component 
from the pion cloud contribution will be somewhat arbitrary. Here the neutron 
with a total zero bulk charge promises an experimental access since one way 
the neutron could acquire a charge distribution is just by its virtual 
dissociation $n \rightarrow p + \pi^-$. This means that the pion cloud should 
be more clearly visible again as a signal at large radii $r \approx 
\lambda_{Compton} = 1/m_{\pi} \approx 1.4\,$fm in the neutron charge 
distribution. 

One could be tempted to look at the neutron rms radius as in the case of the 
proton in first place. However, the mentioned recoil effect causes the magnetic 
moment to contribute to the small electric form factor. It turns out that the 
electric rms radius of the neutron is a subtle interplay between the recoil 
effect and the charge radius proper. We do not want to elaborate this here. 
A summary can be found in Ref.\,\cite{Drechsel:2007sq}. Instead we will show 
the most recent results for the electric FF of the neutron.

Figure\,\ref{fig:6} shows a compilation of all data including the results from 
the MIT Bates measurement with the BLAST detector at the South Hall Ring at small 
$Q^2$~\cite{Geis:2008ha} and the measurement at the Jefferson Laboratory (JLab) 
by the Hall A Collaboration at large $Q^2$~\cite{Riordan:2010xx}. All these 
measurements use the dependence of the cross sections of polarized targets or 
recoil polarizations of the ejected nucleons for polarized electrons from 
the FFs. The curve shows a fit of the phenomenological model of Friedrich and 
Walcher (FW)~\cite{Friedrich:2003iz}. 

\begin{figure}[ht]
\begin{center}
\includegraphics[width=\columnwidth]{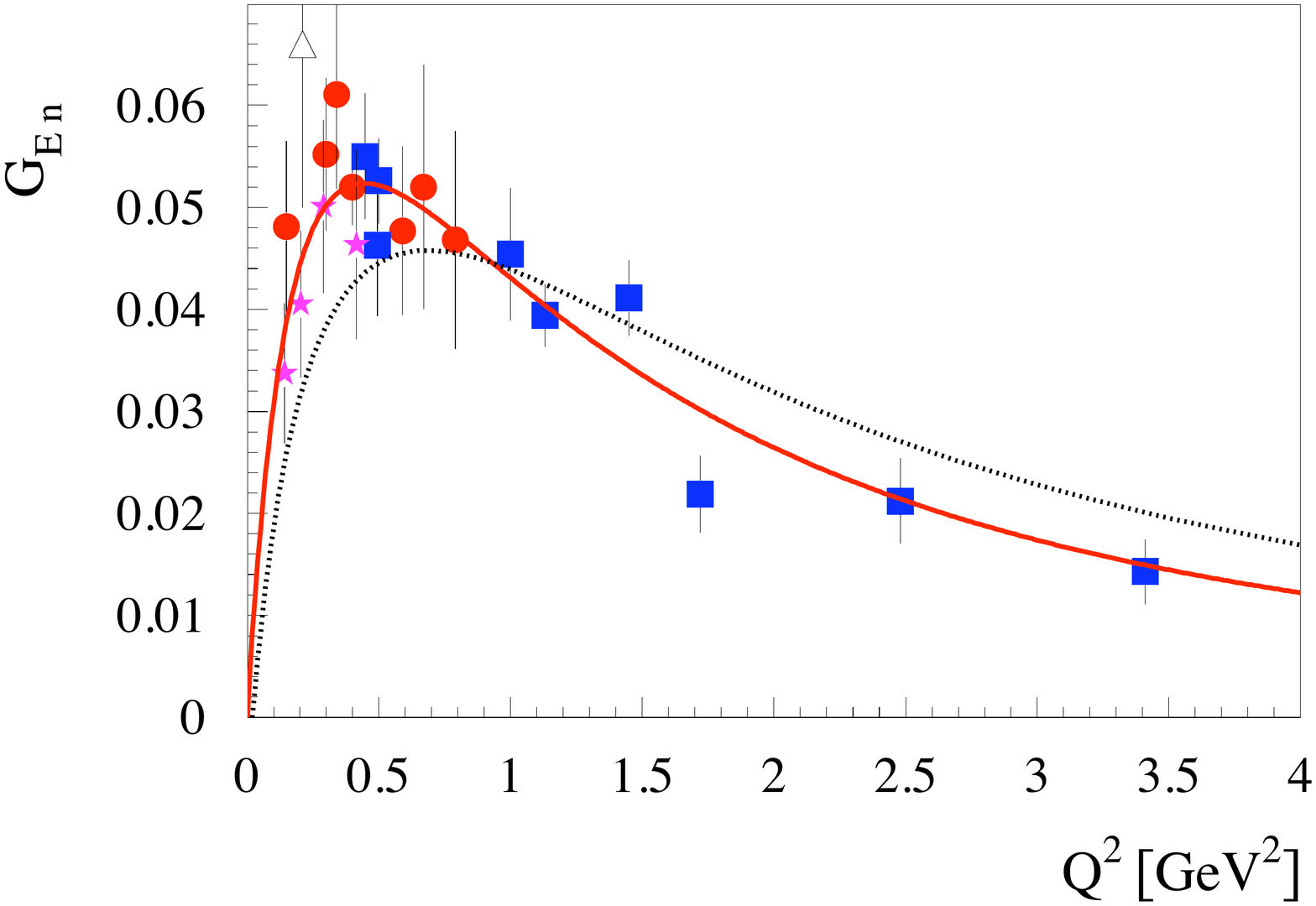}
\end{center}
\caption{A compilation of the recent data for the neutron electric FF, 
$G_{En}$, obtained at NIKHEF (triangle), Bates MIT (stars), JLab (squares), 
and MAMI Mainz (circles), with the quasi free 
$\vec{e}\,d \rightarrow p\,\vec{n}\,e$ and the 
$\vec{e}\,\protect\overrightarrow{^3He} \rightarrow pp\,n\,e$ reactions. 
Either the polarization of the recoiling neutron $\vec{n}$ is measured or 
the target $\protect\overrightarrow{^3He}$, and with it the neutron, is 
polarized. The  protons $p$ and $pp$ are unobserved, respectively. The 
curves correspond with a new fit of a phenomenological model (solid red 
curve)~\cite{Friedrich:2003iz}, and of a Generalized Parton Distribution 
parametrization (dotted black curve)~\cite{Guidal:2004nd}.}
\label{fig:6}
\end{figure}

As already mentioned, the phenomenological FW model tried a coherent fit of 
a smooth bulk curve with a superimposed bump for the pion cloud. This fit 
of the before 2003 data showed indeed a $2\,\sigma$ contribution above the 
smooth curve causing a shift of the negative charge to radii around 1.5\,fm.
The inclusion of the MIT~\cite{Geis:2008ha} and JLab data~\cite{Riordan:2010xx} 
in the fit makes the bump disappear.
At small $Q^2$ the neutron FFs cannot be reliably determined due to the 
model dependence of the extraction of the form factor from the scattering of 
the polarized electrons from unavoidably bound neutrons in deuterium or $^3$He 
targets. Only new even more precise measurements will be able to improve the 
situation. However, it will be mandatory to further study the reaction mechanism 
experimentally in order to check the theoretical corrections.

The scope of this article is on ``long range structure'' meaning that one wants 
to discuss the idea of the spatial distribution of the constituents of the 
nucleon. In the rest frame the three-dimensional charge distribution of a 
spherically symmetric non-relativistic system is obtained as~:
\begin{eqnarray}
\rho_{3d}(r) &=& \int_{0}^{+ \infty} \frac{d k}{2 \pi^2} \, k^2 \, j_0(r k) \, 
\tilde \rho(k),
\label{eq:3dimstatic}
\end{eqnarray}
where $\tilde \rho(k)$ is an intrinsic FF. 
As already pointed out the relativistic effects do not allow for a simple 
interpretation of the electric and magnetic FFs in terms of charge and 
magnetic density distributions in the rest frame system. However, there is 
special reference system, the Breit or brick-wall system, defined by having no 
energy transfer $\nu$ to the nucleon, in which the charge operator for a 
non-relativistic (static) system is only expressed through the electric FF $G_E$. 
In this system the four-momentum transfer squared becomes 
$q^2 = \nu^2 - \vec{q}^{\, 2} = -\vec{q}^{\, 2}$. 
However, the rest frame systems (laboratory systems) for different $q^2$ move 
with different velocities with respect to the Breit system. For a relativistic 
reference system, to relate its  intrinsic FF $\tilde \rho(k)$ and density to 
the Breit frame in which the system of mass $M$ moves with velocity 
$v = \sqrt{\tau / (1 + \tau)}$, requires a Lorentz 
boost relating $k^2 = Q^2 / (1 + \tau)$. This relation shows that for 
$Q^2 \to \infty$, there is a limiting largest intrinsic wave vector 
$k \to 2 M = 2 \pi / \lambda_{lim}$. In the rest frame, no information 
can be obtained on distance scales smaller than this wavelength due to 
relativistic position fluctuations (known as the {\it Zitterbewegung}). For a 
non-relativistic system as $^{16}O$, displayed in Fig.\,\ref{fig:2}, 
$\lambda_{lim}$ is below 0.04\,fm, whereas for the nucleon, it corresponds with 
$\lambda_{lim} \simeq 0.66$\,fm. Extracting the density for a relativistic system 
as the nucleon, therefore requires a prescription in order to relate the 
intrinsic FFs $\tilde \rho(k)$ in Eq.\,(\ref{eq:3dimstatic}) to the 
experimentally measured FFs. As an example, in Ref.\,\cite{Kelly:2002if}, 
the prescription $\tilde \rho(k) = (1 + \tau)^n G_E(Q^2)$ motivated by simple 
models of nucleon structure was used and parameter values $n = 0, 1$, and 2 
were investigated. To see the transitional region from the distance scales 
where relativistic position fluctuations hamper our extraction of rest frame 
densities to distances where the concepts of a non-relativistic many-body 
system can be approximately applied, we visualize the charge density in 
Fig.\,\ref{fig:7}. It depicts the Fourier transform Eq.\,(\ref{eq:3dimstatic}) 
using the fit in Fig.\,\ref{fig:6} (solid red curve). 
One notices a negative charge density at distances around and larger than 
1\,fm. With all caveats we may interpret the negative charge as a ``pion cloud'' 
in the nonrelativistic limit since it extends beyond the confinement radius 
of about 0.8\,fm.

\begin{figure}[h]
\begin{center}
\includegraphics[width=0.95\columnwidth]{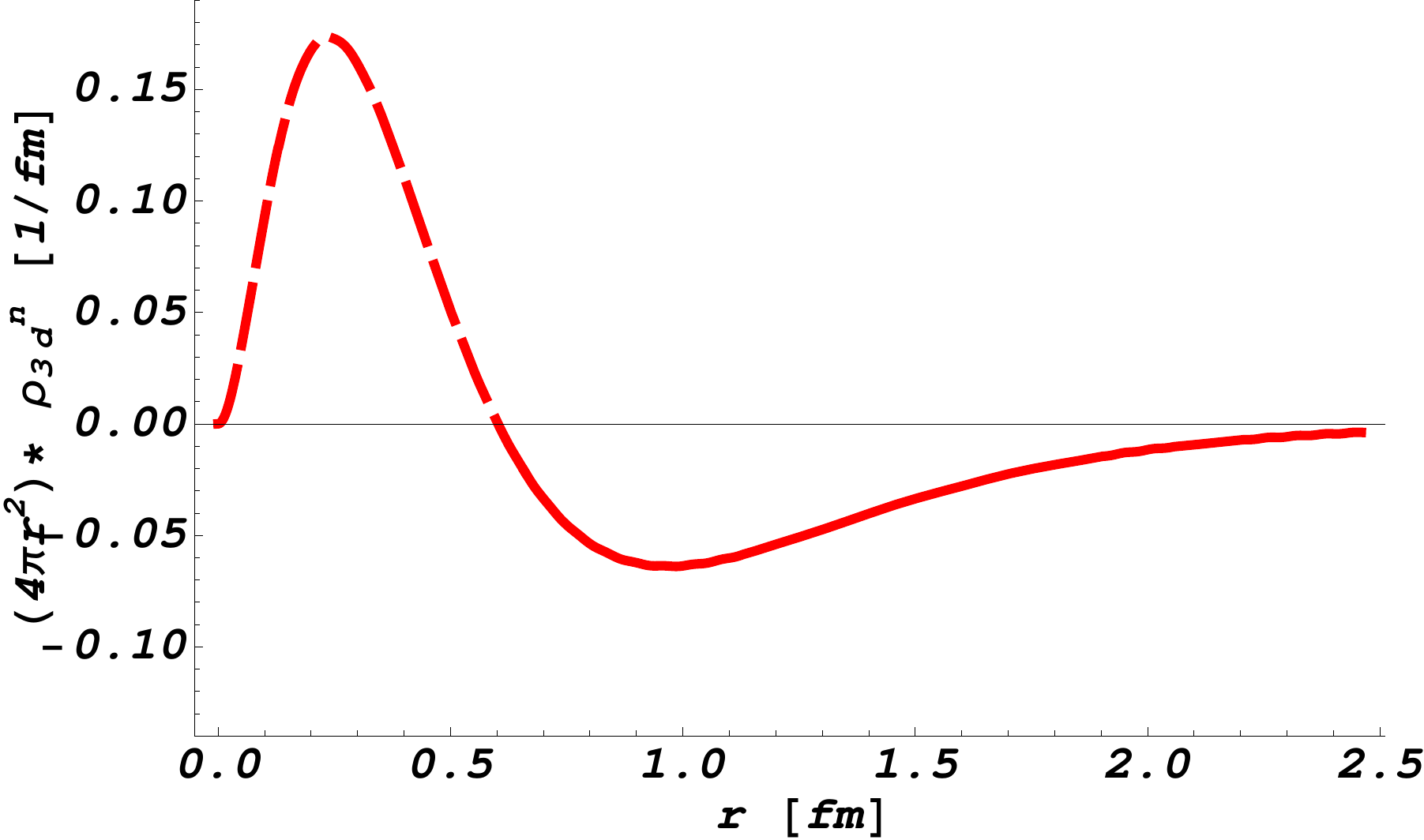}
\end{center}
\caption{Charge distribution of the neutron as derived from the Fourier 
 transform of the $G_{En}$ fit (solid red curve in Fig.\,\ref{fig:6}). 
 The dashed part of the curve is for $r < \lambda_{lim}= 2 \pi / (2 M)$, 
 where one is  intrinsically limited to resolve the density due to the 
 {\it Zitterbewegung} of the nucleon.}
\label{fig:7}
\end{figure}


\section{\sf Relativistic picture} \label{sec:3}

We now turn to the relativistic picture and see how it does complicate matters, 
however, for the benefit of a deeper insight. As already mentioned both, the size 
and the shape of an object, are not relativistically invariant quantities: 
observers in different frames will infer different magnitudes for these 
quantities. Furthermore when special relativity is written in a covariant 
formulation, the density appears as the time component (zero component) 
of a four-current  density $J^{\mu}=(\rho,\textbf{J})$ (in units in which the 
speed of light $c = 1$).

\begin{figure}[h]
\begin{center}
\includegraphics[width =3.5cm]{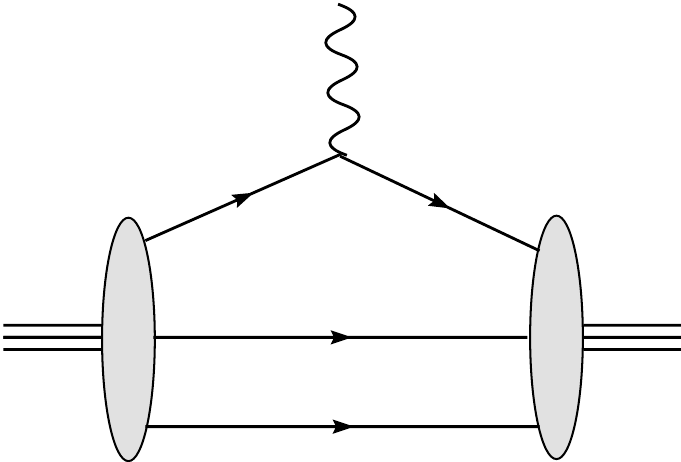}\\
\includegraphics[width =3.5cm]{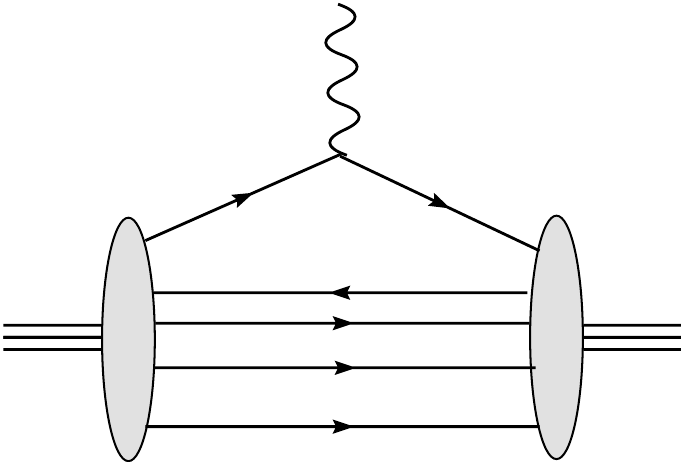}\\
\includegraphics[width =3.5cm]{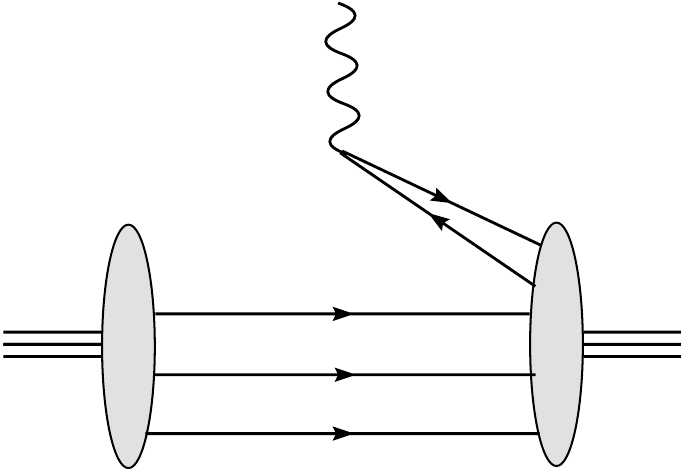}
\caption{Coupling of a space-like photon to a relativistic
many-body system, such as the proton. Top panel (a) : diagonal
transition where the photon couples to a quark in the leading $3q$
Fock component of the proton. Middle panel (b) : diagonal transition
where the photon couples to a quark in a higher Fock  component
(here $4q \bar q$) of the proton. Lower panel (c) : process where the
photon creates a $q \bar q$ pair leading to a non-diagonal
transition between an initial $3q$ state and a final $4q\,\bar{q}$ state. }
\label{fig:overlap}
\end{center}
\end{figure}

Besides the relativistic kinematic effects, as e.g. the length contraction,
the concept of size and shape in relativistic quantum systems, such as hadrons, 
is also profoundly modified as the number of degrees of freedom is not fixed 
anymore. In relativistic quantum mechanics the number of constituents of a 
system is not constant as a result of virtual pair production. We consider as 
an example a hadron such as the proton which is probed by a space-like virtual 
photon, as shown in Fig.\,\ref{fig:overlap}. A relativistic bound state as the 
proton is made up of almost massless quarks. Its three valence quarks making 
up for the proton quantum numbers, constitute only a few percent of the total 
proton mass. In such a system, the  wave function contains, besides the three 
valence quark Fock component  $| qqq \rangle$,  also components where additional 
$q \bar q$ pairs, the so-called sea-quarks, and (transverse) gluons $g$ 
contribute leading to an infinite tower of $ |qqq q \bar q \rangle$, $ |qqq g 
\rangle$, ... components. When probing such a system using electron scattering, 
the exchanged virtual photon will couple to both kind of quarks, valence and 
sea, as shown in Figs.\,\ref{fig:overlap} (a) and (b). In addition, the virtual 
photon, can also split into a $q \bar q$ pair, leading to a transition from a 
$3q$ state in the initial wave function to a $4q \bar q$ state in the final wave 
function, as depicted in Fig.\,\ref{fig:overlap} (c). Such processes representing 
non-diagonal overlaps between initial and final wave functions are not positive 
definite and do not allow for a simple probability interpretation of the density 
$\rho$ anymore. Only the processes shown in Figs.\,\ref{fig:overlap} (a) and (b), 
with the same initial and final wave function yield a positive definite particle 
density allowing for a probability interpretation.

This relativistic dynamic effect of pair creation or annihilation
fundamentally hampers the interpretation of density and
any discussion of size and shape of a relativistic quantum system.
Therefore, an interpretation in terms of the concept of a density 
requires suppressing the contributions shown in Fig.\,\ref{fig:overlap} 
(c). This is possible when viewing the hadron from a light front reference 
frame allowing for a description of the hadron state by an infinite tower 
of light-front wave functions~\cite{Brodsky:1997de}. Consider the 
electromagnetic (e.m.) transition from an initial hadron (with four-momentum 
$p$) to a final hadron (with four-momentum $p^\prime$) viewed from a
light-front moving towards the hadron. Equivalently, this corresponds to 
a frame where the hadrons have a large momentum-component along the 
$z$-axis chosen along the direction of the hadrons average momentum 
$P = (p + p^\prime)/2$. One then defines the light-front plus (+)\,component 
by $a^+ \equiv a^0 + a^3$, in a general four-vector $a^\mu$, 
which is always a positive quantity for both quark or anti-quark four-momenta 
in the hadron. 
When we now view the hadron in a so-called Drell-Yan frame~\cite{Drell:1969km}, 
where the virtual photon four-momentum $q$ satisfies $q^+ = 0$, energy-momentum 
conservation will forbid processes in which this virtual photon splits into a 
$q \bar q$ pair. Such a choice is possible for a space-like virtual photon, and 
its four-momentum or ``virtuality'' is then given by  
$q^2 = - {\vec q_\perp}^{\, 2} \equiv - Q^2 < 0$, where $\vec q_\perp$ is the 
transverse photon momentum lying in the $xy$-plane.
In such a frame, the virtual photon only couples to forward moving partons, 
i.e. only processes such as in Fig.\,\ref{fig:overlap} (a) and (b) are allowed. 
We can then define a proper density operator through the + component of the 
four-current by $J^+ = J^0 + J^3$~\cite{Susskind:1968zz}. For quarks it is given 
by
\begin{equation}
J^+ = \bar q \gamma^+ q = 2 q_+^\dagger q_+^{\phantom{\dagger}}, \quad 
\mathrm{with} \quad q_+ \equiv \frac{1}{4} \gamma^- \gamma^+
q, \label{eq:reloperator}
\end{equation}
where we introduced the $q_+$ fields through a field redefinition from the initial 
quark fields $q$ involving the $\pm$ components of the Dirac gamma matrices.  
The relativistic density operator $J^+$, as defined in Eq.\,(\ref{eq:reloperator}), 
is a positive definite quantity. For systems consisting of e.g. light $u$ and $d$ 
quarks, multiplying this current with the quark charges yields a quark charge
density operator given by $J^+(0) = +2/3 \, \bar u(0) \gamma^+ u(0) -
1/3 \, \bar d(0) \gamma^+(0) d(0)$. Using this charge density
operator, one can then define quark (transverse) charge densities in a
hadron as~\cite{Burkardt:2000za, Miller:2007uy}:
\begin{multline}
\label{eq:dens1}
\rho_0(b) \equiv \int \frac{d^2 \vec q_\perp}{(2
\pi)^2} \, e^{- i \, \vec q_\perp \cdot \vec b} \, \frac{1}{2 P^+} 
\\
\times \langle P^+, \frac{\vec q_\perp}{2},
+\frac{1}{2} \,|\, J^+(0) \,|\, P^+, -\frac{\vec q_\perp}{2}, +\frac{1}{2}
\rangle,
\end{multline}
where the hadron is in a state of definite (light-front) helicity.
In the two-dimensional Fourier transform of Eq.\,(\ref{eq:dens1}), the 
two-dimensional vector $\vec b$ denotes the quark position in the $xy$-plane 
relative to the position of the transverse centre-of-momentum of the hadron. 
It represents the position variable conjugate to the hadron relative transverse 
momentum which equals just the photon momentum $\vec q_\perp$.

The quantity $\rho_0(b)$ has the interpretation of the two-dimensional 
unpolarized quark charge density at a distance $b = | \vec b|$ from the origin of 
the transverse c.m. system of the hadron. In the light-front 
frame, it corresponds to the projection of the charge density in the hadron along 
the line-of-sight. It is important to mind this difference to the interpretation in 
the non-relativistic case.

The quark charge density in Eq.\,(\ref{eq:dens1}) does not fully describe the e.m. 
structure of the hadron, because we know  
that there are two independent e.m. FFs describing the structure of the 
nucleon. In general, a particle of spin $S$ is described by 
$(2 S + 1)$ e.m. moments. In order to fully describe the relativistic structure of 
a hadron one needs to consider additionally the charge densities in a transversely 
polarized hadron state yielding a transverse charge distribution $\rho_{T \, s_\perp}$.
We denote the transverse polarization direction by 
$\vec S_\perp = \cos \phi_S \,\hat e_x + \sin \phi_S \,\hat e_y$.
The transverse charge densities can then be defined through matrix elements of the 
density operator $J^+$ in eigenstates of transverse spin 
as~\cite{Carlson:2007xd,Carlson:2008zc,Lorce:2009bs}:
\begin{multline}
\label{eq:dens2}
\rho_{T \, s_\perp} (\vec b) \equiv \int \frac{d^2 \vec
q_\perp}{(2 \pi)^2} \, e^{- i \, \vec q_\perp \cdot \vec b} \,
\frac{1}{2 P^+} \\ \times
\langle P^+, \frac{\vec q_\perp}{2},
s_\perp | J^+ | P^+, \frac{- \vec q_\perp}{2}, s_\perp  \rangle,
\end{multline}
where $s_\perp$ is the hadron spin projection along the direction of $\vec S_\perp$. 
Whereas the density $\rho_\lambda$ for a hadron in a state of definite helicity is 
circular symmetric for all spins, the density $\rho_{T \, s_\perp}$ depends also on 
the orientation of the position vector $\vec b$, relative to the transverse spin 
vector $\vec S_\perp$. Therefore, it contains the information on the hadron shape, 
again projected on the plane perpendicular to the line-of-sight.  

The light-front wave functions and light-front densities discussed above are defined 
at equal light-front time ($x^+ = 0$) of their constituents. When constituents move 
non-relativistically, it does not make a difference whether they are observed at 
equal time ($t = 0$) or equal light-front time ($x^+ = 0$), since the constituents 
can only move a negligible small distance during the small time interval that a 
light-ray needs to connect them. This is not the case, however, for bound systems 
of relativistic constituents such as hadrons, or even when considering  bound systems
of non-relativistic constituents if the reference system moves relativistically. 
Considering the latter, one may ask, for example, how the equal-time wave functions 
of bound state systems such as the hydrogen atom or positronium, which are 
non-relativistic bound states in QED at leading order in the fine structure constant 
$\alpha$, transform when they move relativistically 
\cite{Jarvinen:2004pi, Hoyer:2009ep}. The case of positronium has been studied in 
Ref.\,\cite{Jarvinen:2004pi}, by using the exact Bethe-Salpeter equation. It was 
found  that the equal-time wave function of its leading $| e^+ e^- \rangle $ Fock 
state contracts as expected from classical relativity. However, calculating the 
bound state in relativistic motion also necessitates the inclusion of an 
$| e^+ e^- \gamma \rangle $ component in the wave function, whose probability is 
of order $O(\alpha)$. Unlike classical transformation laws, it was found that this 
photon amplitude, which can be considered as a quantum fluctuation, does not 
contract~\cite{Jarvinen:2004pi}. Therefore, the Fock components have a more complex 
shape change than just a flattening of their distributions along the direction of 
the boost. Only very recently, relativistic bound states, such as $q \bar q$ states 
in QCD have also been studied at lowest order in $\hbar$~\cite{Hoyer:2009ep}. It was 
found that the resulting equal-time wave functions have unique Lorentz transformation
properties, ensuring the correct dependence of the bound state energy on the 
center-of-mass momentum. A full understanding of the boost properties of bound state 
wave functions would allow to relate rest frame densities with light-front densities. 
For a relativistic many body system as a nucleon, for which this problem still needs 
to be fully solved theoretically, light-front densities provide at present a 
consistent theoretical method to image the charge density of quarks. Such imaging is 
analogous to a flash photograph where different parts of the exposed object are 
reached by the light ray at different times. 

As summarized in the previous section, e.m. FFs of the nucleon are well measured 
experimental quantities. We will, therefore, discuss the relativistic 
spatial shape as derived 
from these FFs. For a nucleon in a state of definite helicity,  
the transverse quark charge density is obtained from 
Eq.\,(\ref{eq:dens1}) by taking the two-dimensional Fourier transform of its 
Dirac FF $F_1 = (G_E + \tau G_M)/(1 + \tau)$ as~\cite{Burkardt:2000za,Miller:2007uy}:
\begin{equation}
\rho_0 (b) = \int_0^\infty \frac{d Q}{2 \pi} Q \, J_0(b \, Q) F_1(Q^2),
\label{eq:ndens1}
\end{equation}
where $J_n$ denotes the cylindrical Bessel function of order $n$.
Note that $\rho_0$ only depends on $b = |\vec b|$.

\begin{figure*}
\begin{center}
\includegraphics[width =6cm]{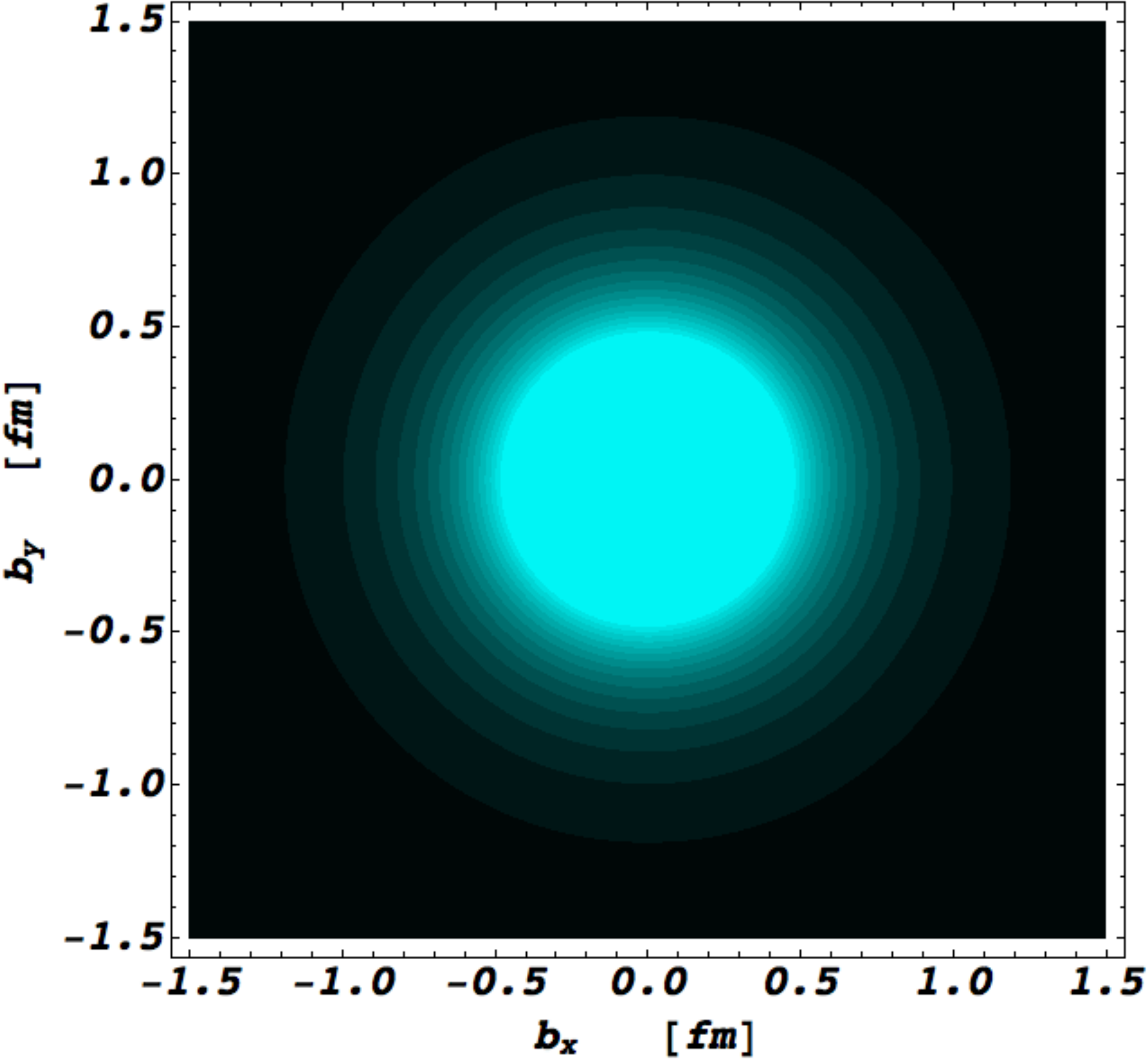}
\includegraphics[width =6cm]{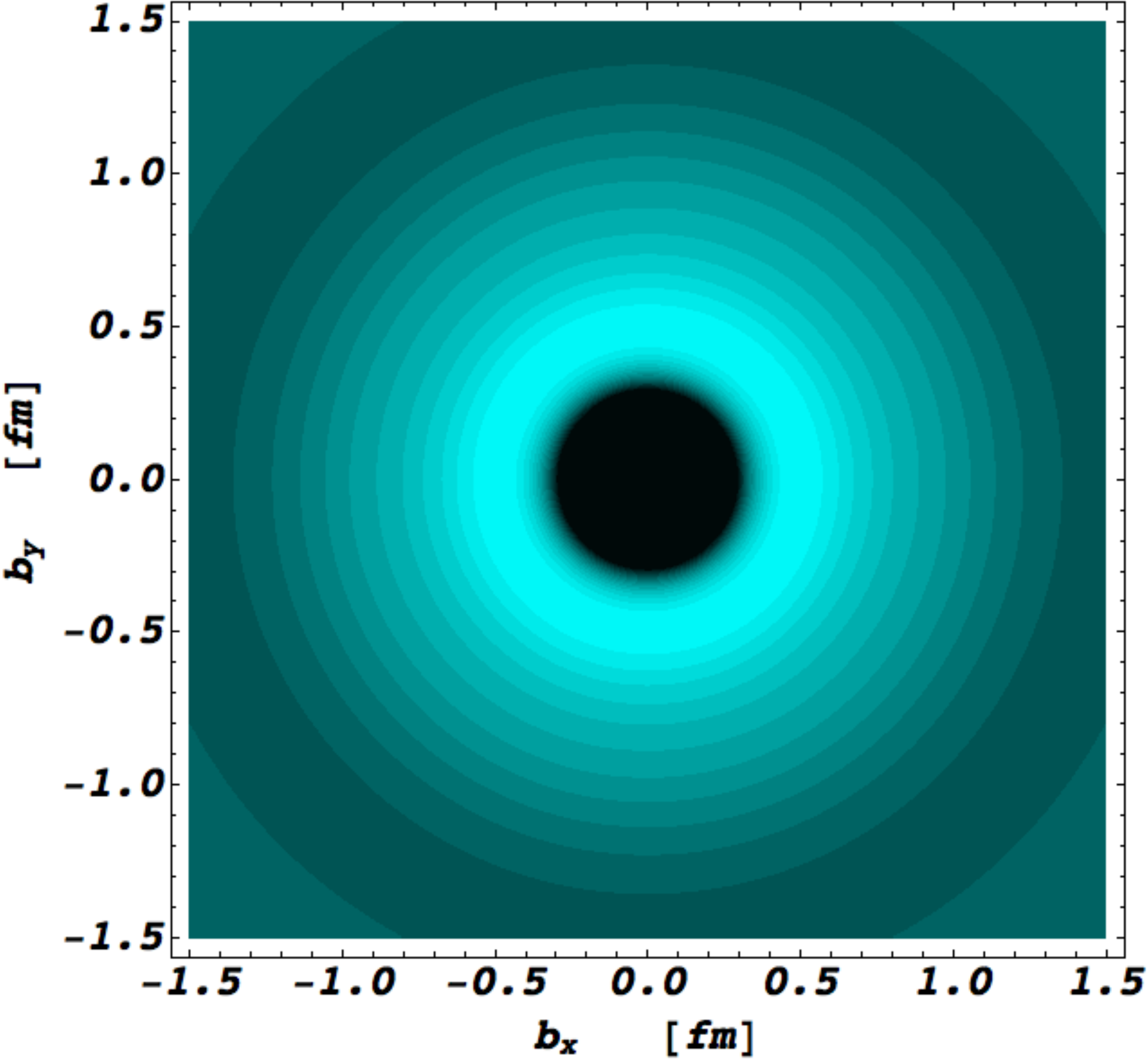}\\
\vspace{0.5cm}
\includegraphics[width=6cm]{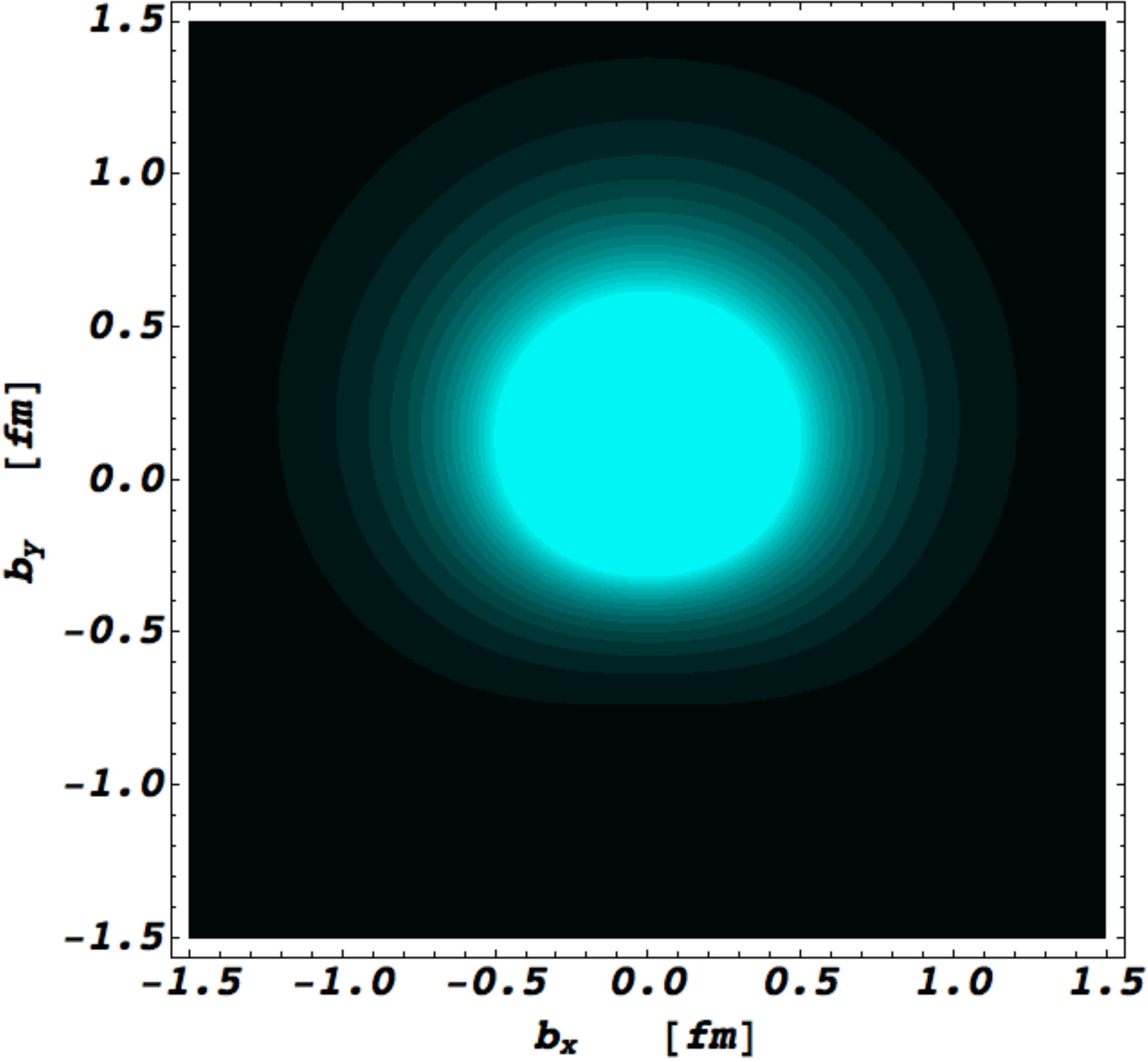}
\includegraphics[width =6cm]{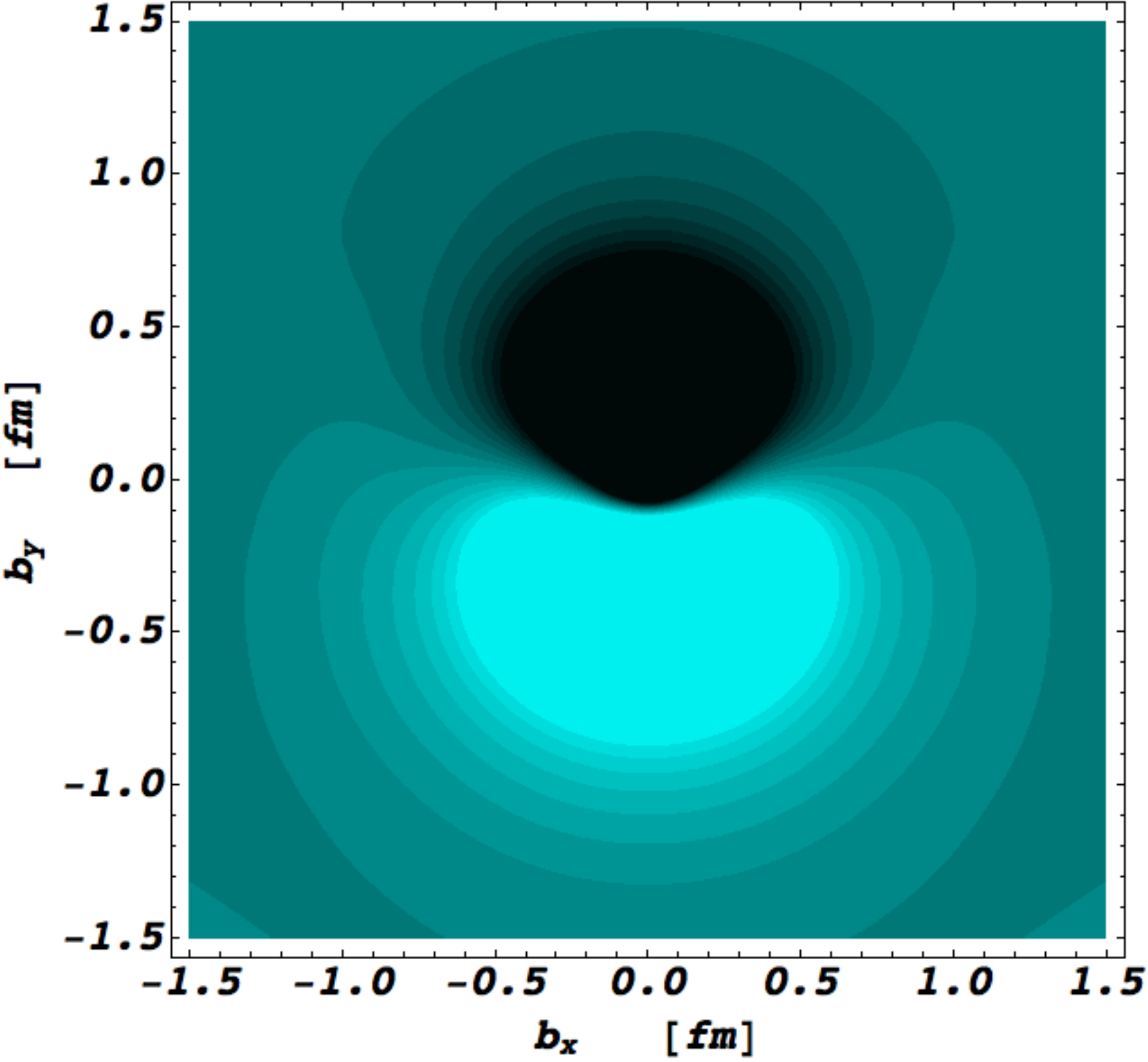}
\caption{Quark transverse charge densities in the {\it proton} (left
panels) and {\it neutron} (right panels). The upper panels show the
density in the transverse plane for a nucleon with definite
helicity. The lower panels for a nucleon polarized along the
$x$-axis. The light (dark) regions correspond with largest
(smallest) values of the density. 
For the proton e.m. FFs, the empirical parametrization of
Arrington {\it et al.}~\cite{Arrington:2007ux} is used. 
For the neutron e.m. FFs, the empirical parametrization of
Bradford {\it et al.}~\cite{Bradford:2006yz} is used.}
\label{fig:protonneutron}
\end{center}
\end{figure*}

\FloatBarrier

On the other hand, the information encoded in the Pauli FF 
$F_2 = (G_M - G_E)/(1 + \tau)$  is connected to a nucleon in a 
transverse spin state. For a nucleon polarized along the positive 
$x$-axis, the transverse spin state can be 
expressed in terms of the light front helicity spinor states by~:
$| s_\perp = + 1/2 \rangle = \left( | \lambda = + 1/2 \rangle
+ e^{i \phi_S } \, | \lambda = - 1/2 \rangle \right) / \sqrt{2}$.
The Fourier transform of the expression given in Eq.\,(\ref{eq:dens2}) 
in a state of transverse spin $s_\perp = +1/2$ then 
yields~\cite{Carlson:2007xd}
\begin{multline}
\rho_{T \frac{1}{2}} (\vec b) = \rho_{\lambda} (b) \\
\quad+ \sin (\phi_b - \phi_S) \,
\int_0^\infty \frac{d Q}{2 \pi} \frac{Q^2}{2 M_N} \, J_1(b \, Q)
F_2(Q^2) .
\label{eq:ndens2}
\end{multline}
The second term, which describes the deviation from the circular symmetric 
unpolarized charge density, depends on the orientation of the transverse 
position vector $\vec b = b ( \cos \phi_b \hat e_x + \sin \phi_b \hat e_y )$, 
relative to the transverse spin direction. 

In Fig.\,\ref{fig:protonneutron}, the transverse charge densities in a 
nucleon, polarized  transversely along the $x$-axis, {\it i.e.} for 
$\phi_S = 0$, are extracted based on the empirical information on the 
nucleon e.m. FFs which, however, does not yet contain the most 
recent data presented in Section\,\ref{sec:2}. For the proton e.m. FFs, 
the empirical parametrization of Ref.\,\cite{Arrington:2007ux} is used,  
whereas for the neutron e.m. FFs, the empirical parametrization 
of Ref.\,\cite{Bradford:2006yz} is taken. One notices from 
Fig.\,\ref{fig:protonneutron} that polarizing the proton along the $x$-axis 
leads to an induced electric dipole moment along the positive $y$-axis 
equivalent to the anomalous magnetic moment $\mu_N$. This field pattern 
due to the induced electric dipole is a consequence of special relativity. 
The nucleon spin along the $x$-axis is the source of a magnetic dipole field 
$\vec B$. An observer moving towards the nucleon with 
velocity $\vec v$ will see an electric dipole field pattern with 
$\vec E^\prime = - \gamma (\vec v \times \vec B)$ giving rise to the observed 
asymmetry.   

For the neutron, one notices its charge density gets displaced significantly 
due to its large negative anomalous magnetic moment  $\mu_N = -1.91$ yielding 
an induced electric dipole moment along the negative $y$-axis. 

To see the impact of the most recent data, presented in Section\,\ref{sec:2}, we 
compare in Fig.\,\ref{fig:9} the new data of Bernauer {\it et al.}~\cite{Ber:2010} 
with the previous parametrization of world data, as performed by Arrington 
{\it et al.}~\cite{Arrington:2007ux}. In order to extract charge densities, one 
requires a form factor parametrization over all values of $Q^2$. Because the 
Bernauer {\it et al.} data only provide a precision measurement of $G_{Ep}$ 
and $G_{Mp}$ for $Q^2  \leq 0.4$\,GeV$^2$, to fully quantify their impact on quark 
charge densities requires a new global analysis combining the previous data with 
these new data. Here we will perform a first estimate of this by using a 
parametrization which smoothly connects the new high precision data at low $Q^2$ 
and the Arrington {\it et al.}~\cite{Arrington:2007ux} parametrization at larger 
$Q^2$. This interpolation function is displayed in Fig.\,\ref{fig:9}, and is used 
to extract the two-dimensional quark charge density in a proton in Fig.\,\ref{fig:10}.
One readily sees that the new high precision data have a direct impact on the 
extracted charge densities at large distances, typically larger than about 1.5\,fm. 
By comparing the extracted density, using the previous fit to world data with the 
new fit, one sees that the new data lead to a significant reduction of the densities 
at distances larger than about 2\,fm.  This is a direct consequence of the flatter 
behavior in $Q^2$, for $Q^2 \leq 0.3$\,GeV$^2$, which the new data display for both 
$G_{Ep}$ and $G_{Mp}$.

\begin{figure}[h]
\begin{center}
\includegraphics[width=0.95\columnwidth]{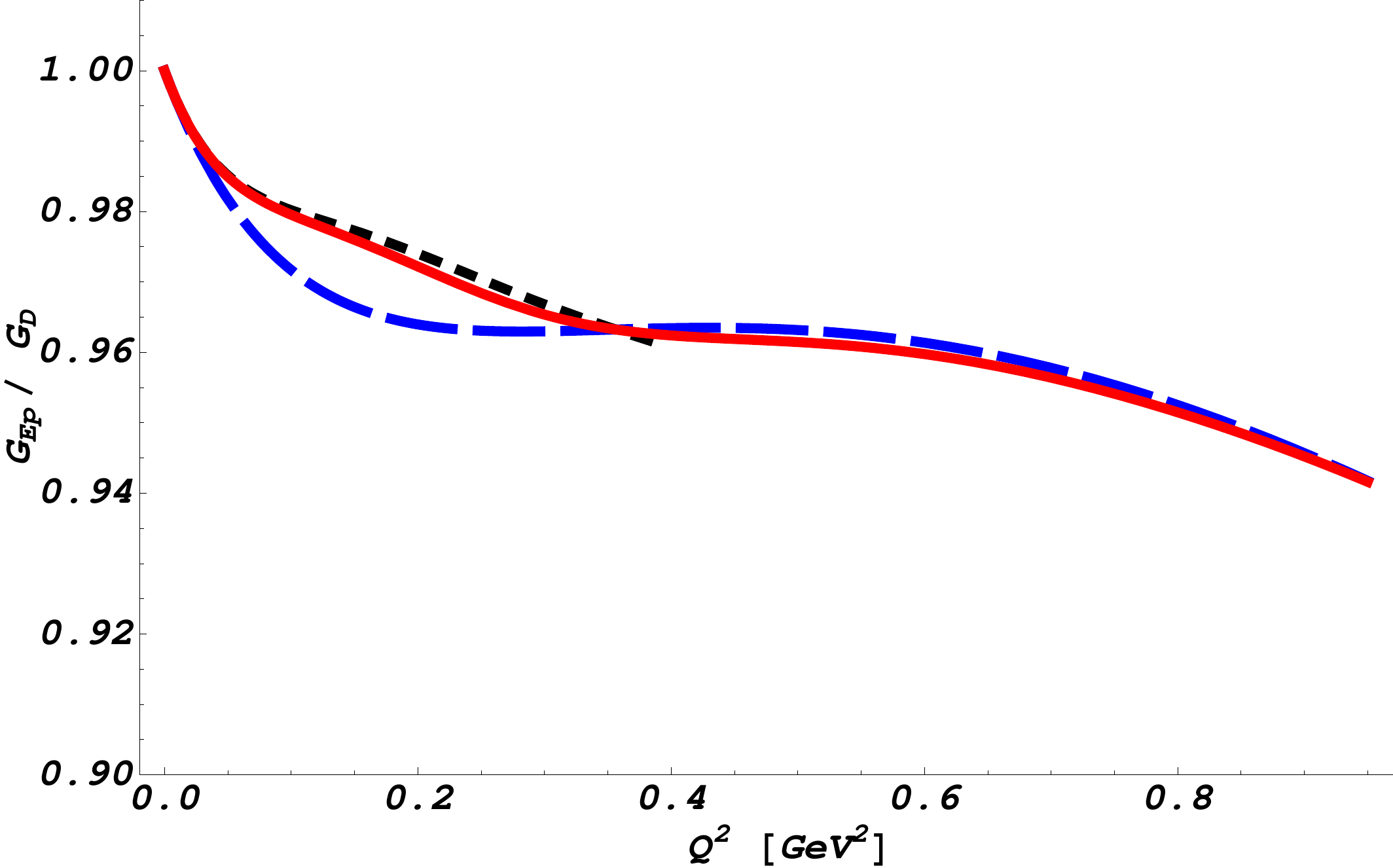}
\includegraphics[width=0.95\columnwidth]{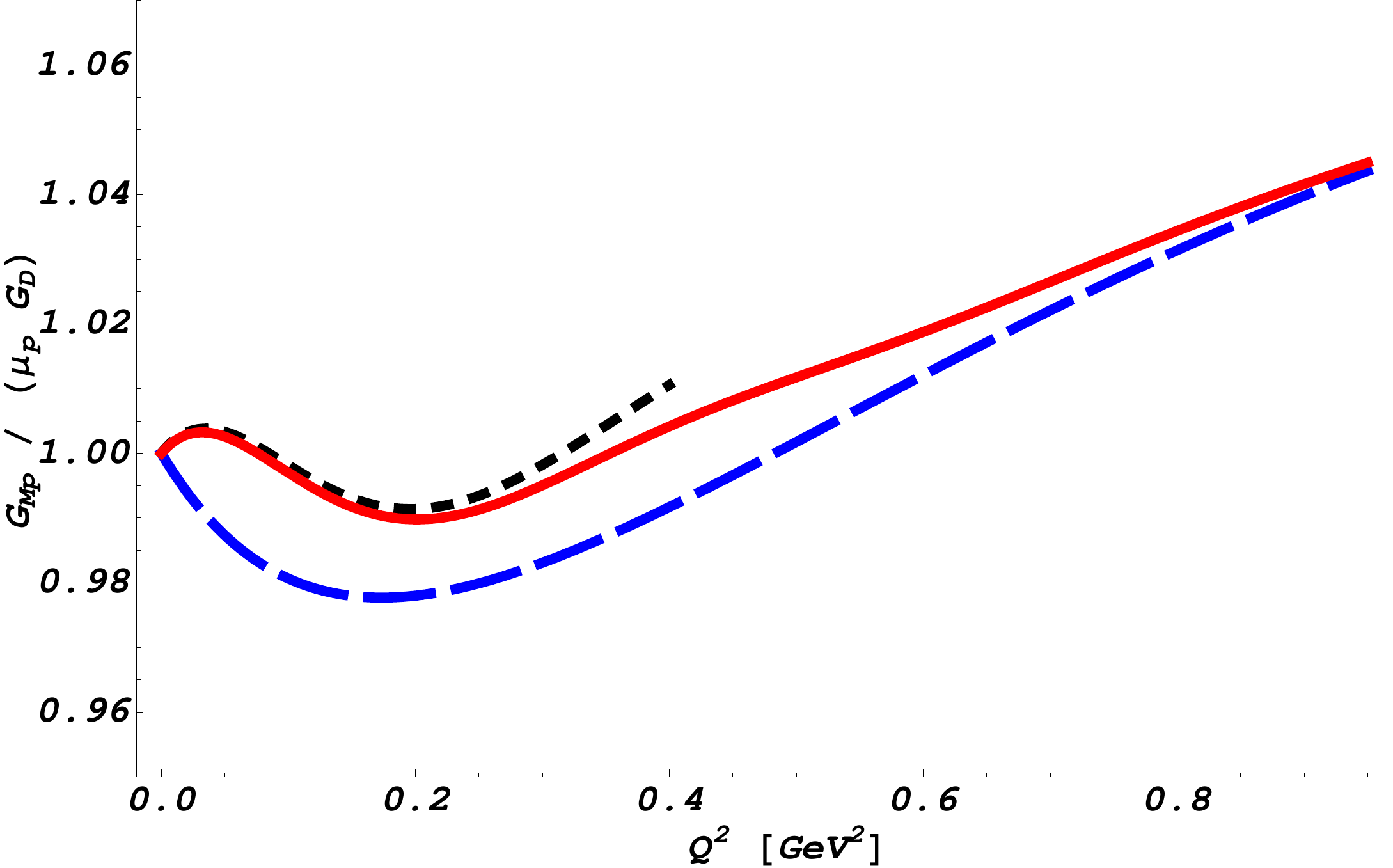}
\end{center}
\caption{Fits to the proton electric (upper panel) and magnetic (lower panel) FFs. 
Blue (dashed) curve: Arrington {\it et al.}~\cite{Arrington:2007ux}, black (dotted) 
curve: 
Bernauer {\it et al.}~\cite{Ber:2010} (only up to $Q^2 = 0.4$~GeV$^2$). 
The red curve presents a smooth connection between the 
Bernauer fit at low $Q^2$ and the Arrington {\it et al.} fit at larger $Q^2$.  }
\label{fig:9}
\end{figure}

\begin{figure}[h]
\begin{center}
\includegraphics[width=\columnwidth]{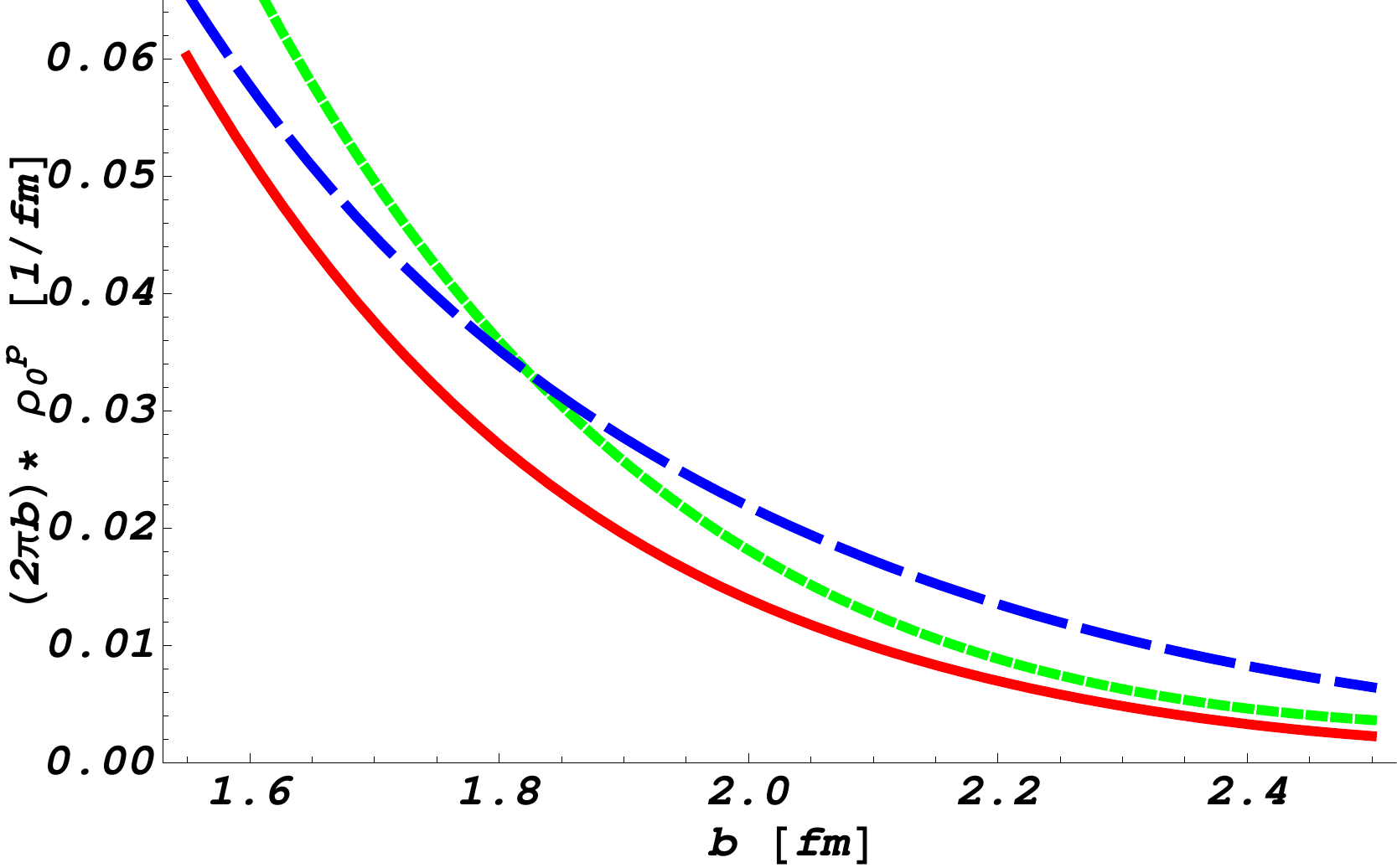}
\end{center}
\caption{Large distance behavior of the unpolarized quark transverse charge density 
in the proton. The dashed blue curve uses the Arrington {\it et al.} 
parametrization~\cite{Arrington:2007ux}. The solid red curve shows the impact of 
recent high precision data at low $Q^2$ by using a smooth connection between the 
Bernauer {\it et al.}~\cite{Ber:2010} fit at low $Q^2$ and the Arrington 
{\it et al.} fit at larger $Q^2$ (see solid red curves in Fig.\,\ref{fig:9}).
For comparison, the solid green curve is the 2-dimensional projection of the 
static charge distribution according to Eq.\,(\ref{eq:2dimstatic}), using the 
interpolating fit for $G_{Ep}$ (red curve in top panel of Fig.\,\ref{fig:9}).}
\label{fig:10}
\end{figure}

In Fig.\,\ref{fig:11}, we show the corresponding large distance behavior of the 
quark charge density in the neutron. The transition between the dashed blue curve 
and the solid red curve in Fig.\,\ref{fig:11} shows the impact of recent precision 
data at low $Q^2$ for the neutron FFs. These lead to a sizable enhancement in the 
extracted densities at distances larger than 1.5\,fm. It is also of interest to 
compare these light-front densities with the static densities as discussed in 
section~2. For a non-relativistic system, one can extract from the 3-dimensional 
static density of Eq.\,(\ref{eq:3dimstatic}), with intrinsic form factor 
$\tilde \rho(k) = G_E(k^2)$, a 2-dimensional static density as~:
\begin{eqnarray}
\rho_{2d}(b) &=& \int_{- \infty}^{+ \infty} d z \, \rho_{3d} (\sqrt{b^2 + z^2}), 
 \nonumber \\
             &=& \int_{0}^{+ \infty} \frac{d Q}{2 \pi} \, Q \, J_0(b Q) \, G_E(Q^2).
\label{eq:2dimstatic}
\end{eqnarray}
One notices that this static 2-dimensional density has the same form as the 
light-front density, see Eq.\,(\ref{eq:ndens1}), with the crucial difference that 
in the static density the Sachs electric FF $G_E$ appears, whereas for a relativistic
system, the proper light-front charge density involves the Dirac FF $F_1$. Since the 
large distance behavior is mostly impacted by the low $Q^2$ data, where $G_E$ is 
dominated by $F_1$,  one expects a qualitatively similar behavior at large distances 
between both pictures. This is illustrated in Figs.\,\ref{fig:10} and \ref{fig:11} 
where the light-front densities (solid red curves) are depicted along with the 
2-dimensional static densities (dotted black curves). One notices that for the 
proton, both densities approach each other at large distances pointing to a large 
tail in the charge distribution. The corresponding picture for the neutron shows 
that both light-front and static densities display a negative charge density for 
distances larger than about 1.6\,fm, which can be associated with a negative pion 
cloud in the outer region of the neutron.

\begin{figure}[h]
\begin{center}
\includegraphics[width=\columnwidth]{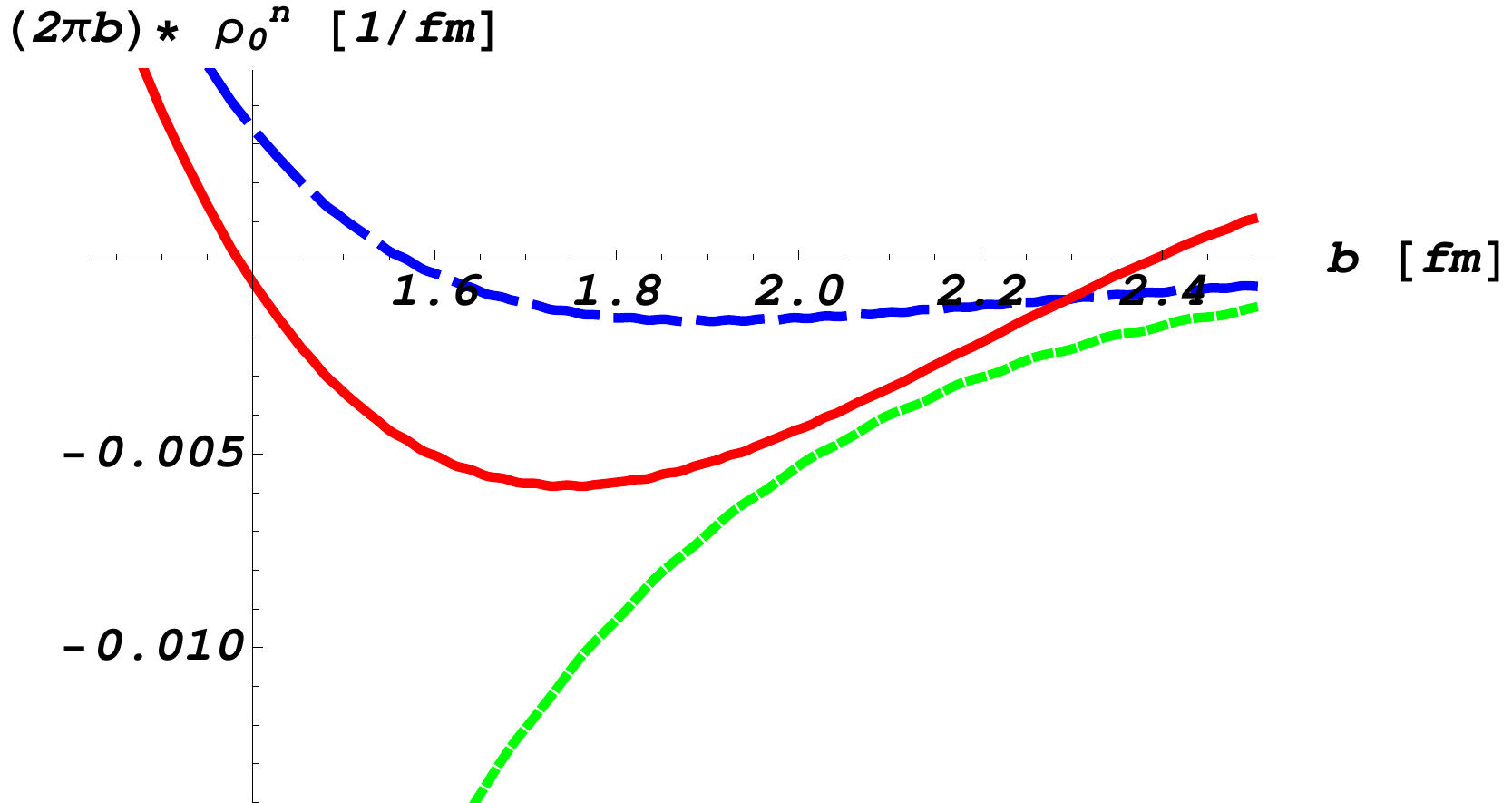}
\end{center}
\caption{Large distance behavior of the unpolarized quark transverse charge density 
in the neutron. The dashed blue curve is the smooth part of the Friedrich-\-Walcher 
parametrization~\cite{Friedrich:2003iz}. The solid red curve is the updated 
Friedrich-\-Walcher parametrization which includes the recent neutron FF data at 
low $Q^2$. For comparison, the solid green curve is the 2-dimensional projection 
of the static charge distribution according to Eq.\,(\ref{eq:2dimstatic}), using the 
most recent fit for $G_{En}$.}
\label{fig:11}
\end{figure}

A combination of the FF data for the proton and neutron allows to perform a quark 
flavor separation and map out the spatial dependence of up and down quarks separately.
 The flavor separated FFs, invoking isospin symmetry, are defined as 
\begin{eqnarray}
F_{1,2 \; u} &=& 2 F_{1,2 \;p} + F_{1,2 \;n}, \nonumber \\
F_{1,2 \; d} &=& F_{1,2 \;p} + 2 F_{1,2 \;n}. 
\end{eqnarray}
For the Pauli FFs, it is convenient to divide out the normalizations at $Q^2 = 0$, 
given by the anomalous magnetic moments $\kappa_u = 2 \kappa_p + \kappa_n$, 
$\kappa_d = \kappa_p + 2 \kappa_n$. 

\begin{figure}[h]
\begin{center}
\includegraphics[width=0.95\columnwidth]{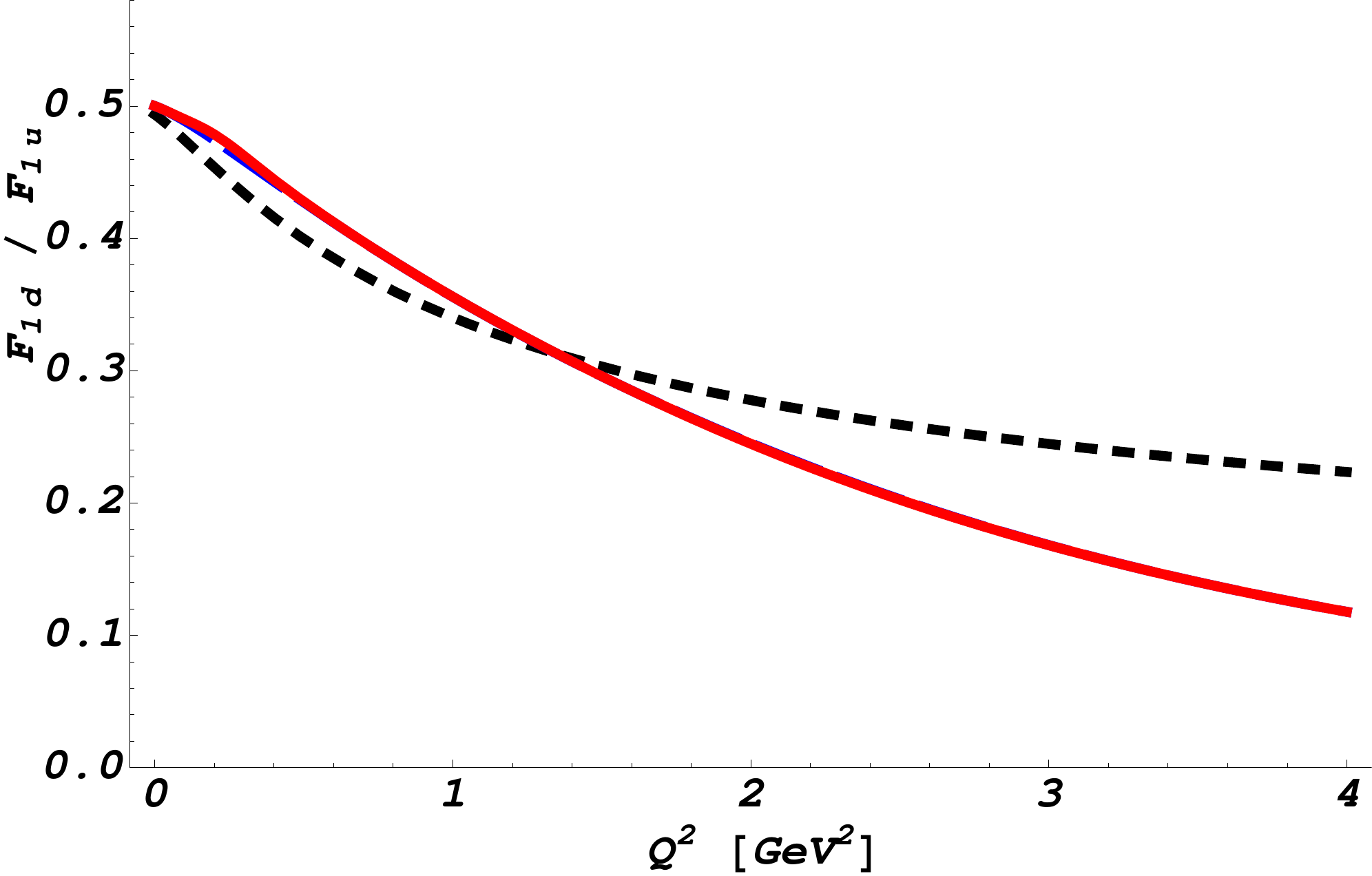}
\includegraphics[width=0.95\columnwidth]{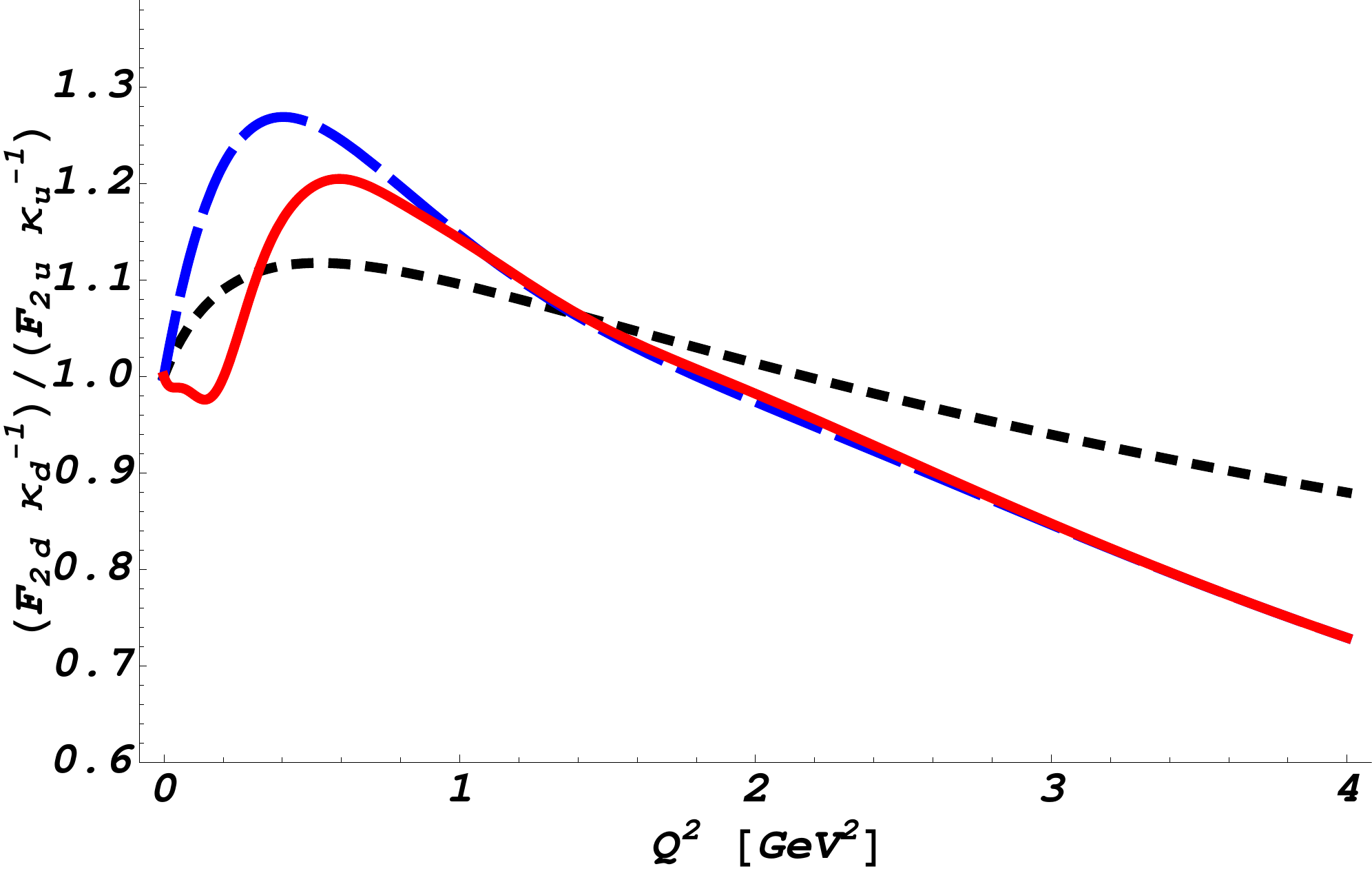}
\end{center}
\caption{Ratio of down/up quark Dirac (upper panel) and Pauli (lower panel) FFs. 
For the Pauli FFs, the anomalous magnetic moments have been divided out. 
The dashed blue curves represent a previous fit to world data, whereas the solid 
red curves shows the impact of recent high precision data at low $Q^2$ as described 
in the text. The dotted black curves represent the results of a Generalized Parton 
Distribution parametrization~\cite{Guidal:2004nd} of up and down quarks. }
\label{fig:12}
\end{figure}

We show the ratio of down/up flavor FFs in Fig.\,\ref{fig:12} and compare 
two empirical parametrizations with the result of a Generalized Parton Distribution 
parametrization of up and down quarks. 
For the empirical parametrizations, the dashed blue curves in Fig.\,\ref{fig:12} 
represent a previous fit to world data, using the  
proton fit of Ref.\,\cite{Arrington:2007ux} (dashed blue curve of  Fig.\,\ref{fig:9}) 
and the dipole type neutron fit of the Friedrich Walcher parametrization 
\cite{Friedrich:2003iz} (solid red curve in Fig.\,\ref{fig:6}). 
The solid red curve in Fig.\,\ref{fig:12} shows the impact of recent data at low 
$Q^2$ both for the proton~\cite{Ber:2010} (solid red curve in Fig.\,\ref{fig:9}), 
as well as the updated Friedrich-\-Walcher parametrization~\cite{Friedrich:2003iz} 
for the neutron electric and magnetic FFs. One clearly sees from Fig.\,\ref{fig:12} 
that the down quark flavor FFs have a faster fall-off than the up quark flavor ones. 
In the phenomenological GPD parametrizations this is encoded through a down-quark 
distribution which drops faster at large momentum fractions $x$ than the up-quark 
distribution. Using Eq.\,(\ref{eq:ndens1}), we can then extract the ratio of up/down 
quark densities in the nucleon, which is displayed in Fig.\,\ref{fig:13}. If the down
and up quarks would have the same spatial distribution in the nucleon, the ratio 
as displayed in Fig.\,\ref{fig:13} would be one. We see however that in the center 
region of the proton, at distances smaller than about 0.5\,fm, down quarks are less 
abundant than up quarks. The down quarks have a much wider distribution and are 
shifted to larger distances, dominating over up quarks between 0.5 to 1.5\,fm. 
At large distances, larger than about 1.5\,fm, one clearly sees the impact of the 
recent data which results in a factor 2 change in the density as compared to previous
fits to world data. Although the contribution of the large distance region to the 
total charge is very small, the new data allow to precisely map out the charge 
densities in the region well beyond the confinement radius, where the charge density 
can in turn be interpreted as a measure of the contribution of the pion cloud. 

\begin{figure}[h]
\begin{center}
\includegraphics[width=0.95\columnwidth]{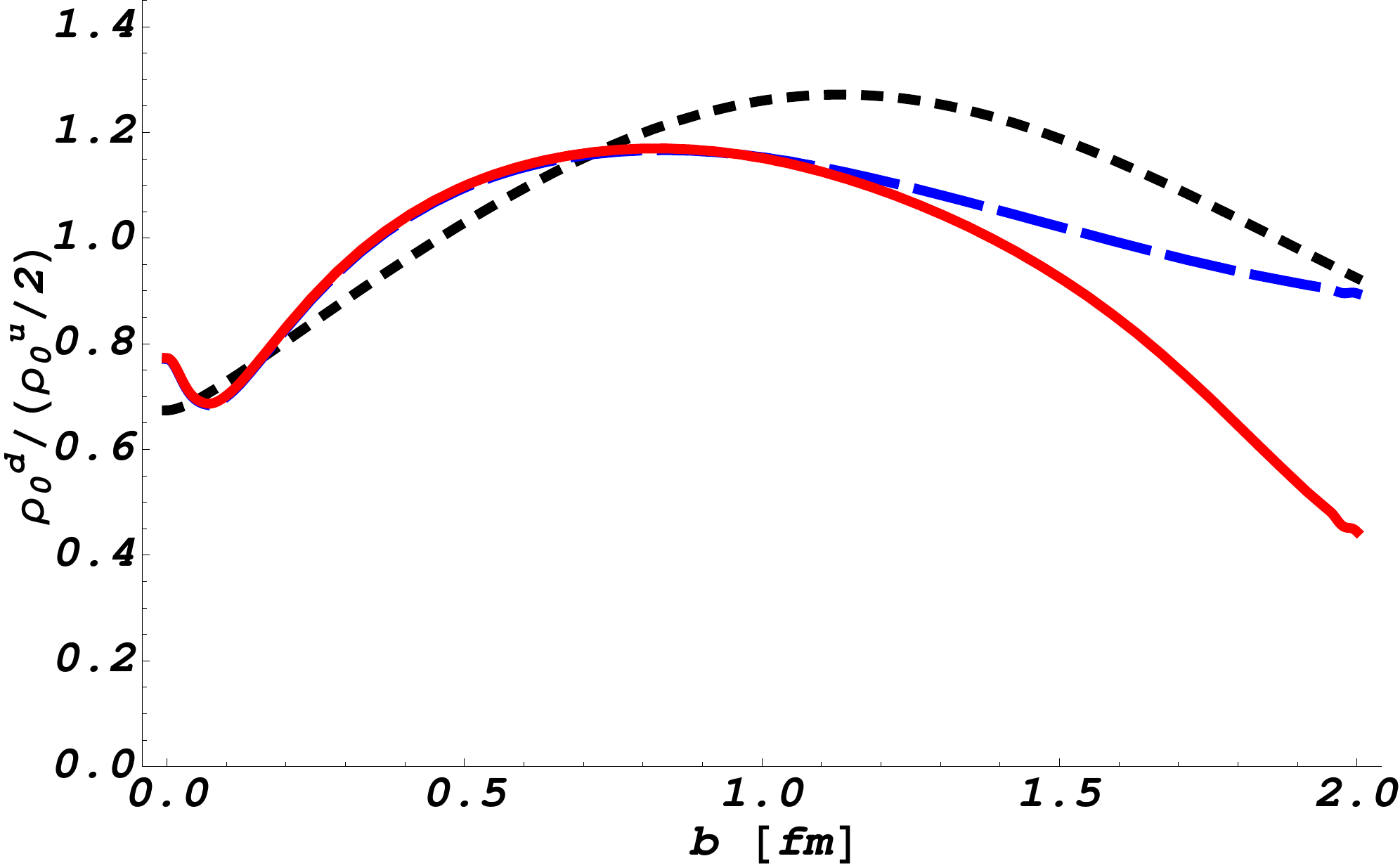}
\end{center}
\caption{Ratio of down over up quark densities in the proton according to the 
parametrizations and curve conventions of Fig.~\ref{fig:12}.  }
\label{fig:13}
\end{figure}


\section{\sf Conclusion}

We have presented two ideas about the long range structure of the nucleon. 
The first is nonrelativistic in terms of a ``bare nucleon'' plus a pion cloud, 
and the second relativistic in terms of quarks and gluons. One may be tempted 
to believe that the second is more fundamental since it uses the elementary 
fields of the standard model of particle physics. The quantum theory of quarks 
and gluons, QCD, describes a very large domain of strong interaction physics 
indeed. However, at sufficiently low energies, hadrons may be described 
by effective field theories formulated in terms of fields with discrete quantum 
numbers. These fields may be viewed as elementary in a certain domain of 
validity, i.e. sufficiently low energies here. One prominent example of such an
effective field is just the pion in Chiral Perturbation Theory emerging as the 
Goldstone Boson of the spontaneous breaking of chiral symmetry of QCD.

In fact, we are not able to devise a quantitative description of the 
nucleon-\-nucleon force in terms of quarks and gluons. On the other hand, 
the meson exchange idea allows for a very precise description of this force. 
Therefore, it may be futile to ask the question which description is more 
correct. Frequently in physics we have to be content with a model allowing 
a description in a limited domain and, following from this, limited predictive 
power.

This sometimes confusing situation is also revealed by the two extreme reference 
frames in which we have considered the structure of the nucleon: the brick-wall 
system implying an infinitely heavy nucleon and the light-front frame implying 
a nucleon moving with approximately the speed of light. 
As we demonstrated in both frames, the long distance structure of the nucleon 
reflects the physics of the pion cloud. With the recent experimental 
and theoretical advances we may have come closer to a common picture. 
However, whether we will ever be able to devise a ``final theory'' 
in terms of the elementary fields is an open question. Actually most physics is 
understood in terms of emergent effective degrees of freedom as the examples of 
condensed matter physics and nuclear physics show overwhelmingly.

\section*{Acknowledgments}
We like to thank J.~Bernauer, J.~Friedrich, K.~de~Jager, V.~Pascalutsa, and 
L.~Tiator for helpful correspondence and discussions.



\begin{thebibliography}{99} 
\itemsep -2pt  
\bibitem{Povh:1986uk}
  B.~Povh and Th.~Walcher,
  Comments Nucl.\ Part.\ Phys.\  {\bf 16} (1986) 85.

\bibitem{Islam:2010}M.~Islam, J. Kaspar, R. Luddy, and A. Prokudin,
  CERN\ Courier, {\bf 49.10} (December 2009) 35.

\bibitem{LanL:1971}
  L.D.~Landau and E.M.~Lifshitz,
  Course of Theoretical Physics, Vol. IV 
  ``Quantum Electrodynamics,'' 
  by W.B.~Berestetskii, E.M.~Lifshitz, and L.P.~Pitaewskii,
  {\it Butterworth-Heinemann, Reed-Elsevier Group,  Oxford (1971).}
 
\bibitem{Lip:1973} 
  H.L.~Lipkin, Quantum Mechanics, New approaches to selected problems.
  {\it North-Holland Publishing Company, Amsterdam (1973).}

\bibitem{Fri:1982}
J.~Friedrich, Physik in unserer Zeit, {\bf 13} (1982) 165.

\bibitem{Neg:1982}    	
J.W.~Negele, Rev.~Mod.~Phys. {\bf 54} (1982) 913.

\bibitem{Ber:2010}
J.~Bernauer, PhD-Thesis, Mainz University 2010, 
  J.~C.~Bernauer {\it et al.}, submitted to Phys.\ Rev.\ Letters,
  arXiv:1007.5076 [nucl-ex].

\bibitem{HydeWright:2004gh}
  C.~E.~Hyde-Wright and K.~de Jager,
  Ann.\ Rev.\ Nucl.\ Part.\ Sci.\  {\bf 54}, 217 (2004). 

\bibitem{Arrington:2006zm}
  J.~Arrington, C.~D.~Roberts and J.~M.~Zanotti,
  J.\ Phys.\ G {\bf 34}, S23 (2007). 

\bibitem{Perdrisat:2006hj}
  C.~F.~Perdrisat, V.~Punjabi and M.~Vanderhaeghen,
  Prog.\ Part.\ Nucl.\ Phys.\  {\bf 59}, 694 (2007). 

\bibitem{Crawford2010:mit}
  C.~Crawford {\it et al.},
  arXiv:1003.0903v3 [nucl-th]

\bibitem{Han:1963}
C.N.~Hand, D.J.~Miller and R.~Wilson, Rev.\ Mod.\ Phys. {\bf 35} (1963) 335.

\bibitem{Sim:1980}
G.G.~Simon {\it et al.} Nucl.\ Phys. {\bf A 333} (1980) 38.

\bibitem{Drechsel:2007sq}
  D.~Drechsel and Th.~Walcher,
  Rev.\ Mod.\ Phys.\  {\bf 80} (2008) 731.

\bibitem{Pohl:2010zz}
  R.~Pohl {\it et al.},
  Nature {\bf 466} (2010) 213.

\bibitem{Friedrich:2003iz}
  J.~Friedrich and T.~Walcher,
  Eur.\ Phys.\ J.\  A {\bf 17} (2003) 607.

\bibitem{Guidal:2004nd}
  M.~Guidal, M.~V.~Polyakov, A.~V.~Radyushkin and M.~Vanderhaeghen,
  Phys.\ Rev.\  D {\bf 72}, 054013 (2005). 

\bibitem{Geis:2008ha}
  E.~Geis {\it et al.}  [BLAST Collaboration],
  Phys.\ Rev.\ Lett.\  {\bf 101} (2008) 042501.
 
\bibitem{Riordan:2010xx}
S. Riordan {\it et al.}, submitted to PRL, arXiv:1008.1738v1 [nucl-ex]
\bibitem{Kelly:2002if}
  J.~J.~Kelly,
  Phys.\ Rev.\  C {\bf 66}, 065203 (2002).

\bibitem{Brodsky:1997de}
  S.~J.~Brodsky, H.~C.~Pauli and S.~S.~Pinsky,
  Phys.\ Rept.\  {\bf 301}, 299 (1998).

\bibitem{Drell:1969km}
  S.~D.~Drell and T.~M.~Yan,
  Phys.\ Rev.\ Lett.\  {\bf 24}, 181 (1970).

\bibitem{Susskind:1968zz}
  L.~Susskind,
  Phys.\ Rev.\  {\bf 165} (1968) 1547.

\bibitem{Burkardt:2000za}
  M.~Burkardt,
  Phys.\ Rev.\  D {\bf 62}, 071503 (2000);
  Int.\ J.\ Mod.\ Phys.\  A {\bf 18}, 173 (2003).

\bibitem{Miller:2007uy}
  G.~A.~Miller,
  Phys.\ Rev.\ Lett.\  {\bf 99}, 112001 (2007).

\bibitem{Carlson:2007xd}
  C.~E.~Carlson and M.~Vanderhaeghen,
  Phys.\ Rev.\ Lett.\  {\bf 100}, 032004 (2008).

\bibitem{Carlson:2008zc}
  C.~E.~Carlson and M.~Vanderhaeghen,
  Eur.\ Phys.\ J.\  A {\bf 41}, 1 (2009).

\bibitem{Lorce:2009bs}
  C.~Lorc\'e,
  Phys.\ Rev.\  D {\bf 79}, 113011 (2009).

\bibitem{Jarvinen:2004pi}
  M.~Jarvinen,
  Phys.\ Rev.\  D {\bf 71}, 085006 (2005). 

\bibitem{Hoyer:2009ep}
  P.~Hoyer,
  arXiv:0909.3045 [hep-ph].

\bibitem{Hoyer:2009sg}
  P.~Hoyer and S.~Kurki,
  Phys.\ Rev.\  D {\bf 81}, 013002 (2010). 

\bibitem{Arrington:2007ux}
  J.~Arrington, W.~Melnitchouk and J.~A.~Tjon,
  Phys.\ Rev.\  C {\bf 76}, 035205 (2007).

\bibitem{Bradford:2006yz}
  R.~Bradford, A.~Bodek, H.~Budd and J.~Arrington,
  Nucl.\ Phys.\ Proc.\ Suppl.\  {\bf 159}, 127 (2006).

\end{thebibliography}
\end{document}